\documentclass[12pt]{article}
\columnwidth\textwidth

\begin{document}

\newcommand{\be}{\begin{equation}}
\newcommand{\ee}{\end{equation}}
\newcommand{\beann}{\begin{eqnarray*}}
\newcommand{\eeann}{\end{eqnarray*}}
\newcommand{\bea}{\begin{eqnarray}}
\newcommand{\eea}{\end{eqnarray}}
\newcommand{\nn}{\nonumber}
\newcommand{\ben}{\begin{enumerate}}
\newcommand{\een}{\end{enumerate}}
\newtheorem{df}{Definition}
\newtheorem{thm}{Theorem}
\newtheorem{lem}{Lemma}
\newtheorem{prop}{Proposition}
\begin{titlepage}

\noindent
\hspace*{11cm} BUTP-96/18 \\
\vspace*{1cm}
\begin{center}
{\LARGE Lagrangian and Hamiltonian Formalism \\[0.5cm] for Discontinuous Fluid 
and Gravitational Field}

\vspace{2cm}

P.~H\'{a}j\'{\i}\v{c}ek \\
Institute for Theoretical Physics \\
University of Bern \\
Sidlerstrasse 5, CH-3012 Bern, Switzerland \\
\vspace*{1cm}

J.~Kijowski \\
Center for Theoretical Physics \\
Polish Academy of Sciences \\
Aleja Lotnik\'{o}w 32/46, 02-668 Warsaw, Poland \\
and \\
Department of Mathematical Methods in Physics \\
University of Warsaw \\
Ul. Ho\.{z}a 74, 00-682 Warsaw, Poland \\
\vspace*{2cm}

July 1997 \\ \vspace*{1cm}

\nopagebreak[4]

\begin{abstract}
  The barotropic ideal fluid with step and $\delta$-function discontinuities
  coupled to Einstein's gravity is studied. The discontinuities represent star
  surfaces and thin shells; only non-intersecting discontinuity hypersurfaces
  are considered.  No symmetry (like eg.\ the spherical symmetry) is assumed.
  The symplectic structure as well as the Lagrangian and the Hamiltonian
  variational principles for the system are written down.  The dynamics is
  described completely by the fluid variables and the metric on the fixed
  background manifold. The Lagrangian and the Hamiltonian are given in two
  forms: the volume form, which is identical to that corresponding to the
  smooth system, but employs distributions, and the surface form, which is a
  sum of volume and surface integrals and employs only smooth variables. The
  surface form is completely four- or three-covariant (unlike the volume
  form). The spacelike surfaces of time foliations can have a cusp at the
  surface of discontinuity. Geometrical meaning of the surface terms in the
  Hamiltonian is given. Some of the constraint functions that result from the
  shell Hamiltonian cannot be smeared so as to become differentiable functions
  on the (unconstrained) phase space. Generalization of the formulas to more
  general fluid is straifgtforward.

\end{abstract}

\end{center}

\end{titlepage}

\section{Introduction}
\label{sec:intro}
Spherically symmetric thin shells or dust stars (like the
Oppenheimer-Snyder one) are popular models used extensively in the study
of a number of phenomena: properties of classical gravitational collapse
\cite{israel-coll}, properties of classical black holes \cite{farrug},
quantum gravitational collapse \cite{H-K-K}, the dynamics of domain
walls in early Universe \cite{guth}, the back reaction in Hawking effect
\cite{K-W}, entropy on black holes \cite{york} or quantum theory of
black holes \cite{berez}, \cite{becken}, to mention just few examples.

The classical dynamics of objects with discontinuities in matter density is
well-understood; it is determined by Einstein's equations, the matter
dynamical equations and some jump conditions at the discontinuity. The jump
conditions for the step-like discontinuity require that there are coordinates
in which the metric is $C^1$ at the discontinuity surface \cite{MTW}; for the
thin shells, they have been first formulated by Dautcourt \cite{dautc};
Dautcourt's equations have been rewritten in a covariant form by Israel
\cite{israel}.

In many investigations, however, a variation principle, or a Hamiltonian is
needed from which this classical dynamics follows. Often, such principles
(suitable eg.\ for spherically symmetric models) are just guessed from the
dynamical equations; some attempts to obtain them from more general
variational principles are \cite{H-L} and \cite{K-W}. Indeed, this is an
interesting problem by itself: how the large number of different
one-degree-of-freedom Hamiltonians scattered in literature is related to the
Einstein-Hilbert action? For our overal picture of the world has to be
self-consistent, even if we indulge in using a number of different models,
each just applicable for a situation under study.

In the present paper, we reformulate the dynamics of gravitation and ideal
discontinuous fluid in the Hamiltonian form. That is, we identify the
canonical variables ($p$'s and $q$'s) and Lagrange multipliers and write down
a Hamiltonian functional of these variables; we show that the constraints and
the canonical equations resulting from this Hamiltonian are equivalent to the
system of Einstein equations and the ideal fluid dynamical equations (plus the
Israel equations in the case of a thin shell).

To identify the suitable symplectic structure and find the variational
formulas, we employ the methods described in detail in \cite{K-T} and their
application to general relativity as given in \cite{JK2}. We will, however,
keep the paper self-contained by motivating and explicitly performing all
relevant derivations.

The model of matter used extensively in this paper is that of the simplest
kind: the barotropic ideal fluid. This can be formulated as a Lagrangian field
theory without any constraints \cite{JK1}. Generalization to ideal fluid with
internal degrees of freedom (such as \cite{brown}) or to any conservative
continuum should be straightforward; in any case, the gravitational parts of
our Lagrangians and Hamiltonians (which represent the solution to the main
problem) have general validity.

In each particular case, the classical dynamics can be obtained from a
variational principle that has the same form as the corresponding variational
principle for a smooth system, if some particular generalized functions are
allowed to describe the matter distribution: the step function for star
boundaries and the $\delta$-function for thin shells (cf.\ \cite{H-L}). This
simplicity is, however, traded for the freedom in the choice of coordinates:
the generalized function approach works only if the metric is $C^1$ for the
step, and $C^0$ for the $\delta$-discontinuity. We transform, therefore, the
Lagrangians and the Hamiltonians to the so-called {\em surface form}
containing only smooth variables; such Lagrangians and Hamiltonians as well as
symplectic forms decompose into sums of volume and surface integrals. The
transformation can best be done in the so-called {\em adapted coordinates};
these are coordinates in which the embedding functions of the surfaces of
discontinuity acquire the simplest possible form. The result, however, is
covariant in the sense that arbitrary smooth coordinates can be chosen inside
of each separated volume (left or right to the discontinuity surface) as well
as along the discontinuity surface itself.

An important trick is used throughout the paper:
we work in coordinate systems which are always adapted to the position of
the discontinuity surface. This way the discontinuity surface may be
considered as a fixed submanifold of the spacetime. Thus, the dynamics
of the star surface or thin shell is not described by the spacetime
coordinates of these objects but by the evolution of the physical fields
like metric of matter fields along the surfaces. Then, for example the
variations and time derivatives of the embedding functions of the
two-surfaces of discontinuity in the three-surfaces of constant time
both vanish identically. Our formulas are written only for one
hypersurface of discontinuity; an extension to arbitrary many
hypersurfaces is easy if they do not intersect each other.

Two interesting problems arise. First, we do not show that the dynamics makes
sense even on-shell. By that, we mean that there is to be a well-posed initial
value problem. One ought to be able to define some nice space of initial data,
consisting of those values of the canonical variables that satisfy some
well-defined set of constraints, jump and fall-off conditions so that a unique
solution to the dynamical equations will exist in a neighbourhood of the
initial surface. In this paper, we shall just assume that the dynamics is all
right. At least in some special cases (like spherical symmetry), the space of
classical solutions is well-known and it is as large as one expects.

The second problem is to show that the Hamiltonian formalism defines a
(regular) constrained system. This means that one can find a phase space
(possibly an extension of ours), a complete set of constraints, and a
Hamiltonian satisfying the following conditions: 1) the constraints and
Hamiltonian must be differentiable functions on the phase space so that their
Poisson brackets are well-defined and 2) the Hamiltonian must be first class
and the constraint set must be split nicely into the first and second class
constraints (Bergmann-Dirac analysis, cf.\ \cite{EN}). Of course, such an
`off-shell' formulation is necessary as a starting point for Dirac
quantization. The difficulty is that some constraints at the shell are not
differentiable functions on the phase space even if they are smeared along the
shell, because the smearing is then only two-dimensional, whereas the
differentiability would require a three-dimensional smearing. Without an
off-sell formulation, the way to quantum theory need not be barred however.
One can try to solve the singular constraints and to substitute the solution
back into the action so that a variational principle results which leads to
equivalent dynamics without the singular constraints \cite{H-T}. A problem
with such a procedure seems to be that equations quickly grow very messy.

The plan of the paper is as follows. Sec.\ 2 is devoted to the step, Sec.\ 3
to the $\delta$-function discontinuities. Sec.\ 2.1 introduces the ideal fluid
model and its dynamics in a fixed spacetime (metric) background. Basic
formulas of the Lagrange and Euler pictures concerning Lagrangians,
Hamiltonians, strees-energy tensors and equations of motion are derived; these
equations apply to both step and $\delta$ function discontinuity. The method
of {\em variation formulas} is presented, which enables us to find the
symplectic structure as well as to generate the equations of motion. The
surface of discontinuity can be moved without problems as far as the metric is
fixed.  In Sec.\ 2.2, the fluid is coupled to the dynamical gravity. Relevant
formulas concerning the variation of the Einstein-Hilbert action are
collected. The surfaces of discontinuity are now fixed. This helps to avoid
some formal problems. The variation formulas for the system are written in
Lagrangian and Hamiltonian form.

In Sec.\ 3.1, an action for the thin shell and dynamical gravity is written
down in the Lagrangian formalism; the shells are fixed and generalized
functions are employed. In Sec.\ 3.2, the adapted coordinates are used to
transform the (Lagrange formalism) action into a sum of volume and surface
integrals disposing of the generalized functions and gaining more coordinate
(gauge) freedom (four-covariance): arbitrary coordinates can be chosen left to
the shell, right to the shell, and along the shell. In Sec.\ 3.3, the
variation of the action in the surface form is calculated and the obtained
dynamical equations are listed; they contain Israel's equation. The variation
formula is derived; this is only three-covariant: the foliation by spacelike
surfaces $t$ = const must be such that the $t$-surfaces are continuous but can
have a cusp at the discontinuity surface; the embedding functions of the
discontinuity two-surface in the $t$-surfaces must be time-independent. Sec.\
3.4 contains a Legendre transformation to a Hamiltonian formalism; the general
form of the Hamiltonian for the system of thin shells and gravity is
presented. In sec.\ 3.5, the explicit functional dependence of the Hamiltonian
on the dynamical variables is written down and the geometrical meaning of the
surface terms in the Hamiltonian is disclosed. In Sec.\ 3.6, the variation of
the Hamiltonian is explicitly calculated so that all canonical equations and
constraints following from the Hamiltonian can be listed. This not only
enables us to check that the Hamiltonian generates the desired dynamics
(including Israel's equation) but also to classify the resulting equations
into `canonical equations' and `constraints.' For example, the six relations
that are equivalent to Israel's equation consist of one super-Hamiltonian
constraint, two supermomentum constraints, two singular constraints (these
cannot be made differentiable by smearing), and one canonical equation. Some,
necessarily preliminary, discussion of the result is given.

\section{Fluid with a step discontinuity}
Our point of departure in this section is the description of relativistic
barotropic perfect fluid as given in Ref.\ \cite{JK1}
(observe that this description is easily extended to any conservative
continuum). We will extend and modify the method so that it allows for
discontinuous matter distributions admitting such situation like a jump of
density at the boundary of a star (a step-function type of discontinuity along
a timelike hypersurface).

\subsection{Fluid in gravitational field}
\subsubsection{The description of the fluid}
\label{sec:lagrang}
The fluid that have just `mechanical' degrees of freedom consists of
identifiable elementary volumes---mass points of the fluid. It can, therefore,
be completely described by specifying the mass and the spacetime coordinates
of each of these mass points. All mass points form the so-called matter space
$Z$, which is a three dimensional manifold; let $z^a$, $a = 1,2,3$, be some
coordinates in $Z$. Let us denote the spacetime by $M$ and let $x^\mu$, $\mu =
0,1,2,3$ be some coordinates in $M$. The state of the fluid can then be
described by a map $\zeta : M \mapsto Z$, in coordinates $z^a(x^\mu)$; the
particle trajectories are then determined by $z^a(x) =$ const. The matter
space $Z$ is equipped with a scalar density $h(z)$, which determines the mole
or particle density of the fluid, so that the number N of particles or moles
in the volume $V_z \subset Z$ is given by N$(V_z) = \int_{V_z}d^3z\,h$. We
assume further that $h$ has a step discontinuity at a two surface $\Sigma_z$
in $Z$, defined by the equation $F(z) = 0$, where $F$ is a smooth function
with non-zero gradient $F_a$. Let $\Sigma := \zeta^{-1}(\Sigma_z)$ be a
timelike three-surface separating $M$ in two open subsets $V^+$ and $V^-$ so
that $h(z(x)) > 0$ for $x \in V^-$ and $h(z(x)) = 0$ for $x \in V^+$. One can
make much more general assumptions (e.g.\ allowing for several matter filled
regions), but this will only complicate the description without requiring any
new method of approach.

The map $\zeta$ and the density $h$ define mole (particle) current $j^\mu$ in
$M$ by
\be
  j^\mu = h\epsilon^{\mu\nu\kappa\lambda} z^1_\nu z^2_\kappa
  z^3_\lambda,
\label{1.1}
\ee
where
\[
  z^a_\mu := \frac{\partial z^a}{\partial x^\mu}.
\]
$j^\mu(x)$ is discontinuous at $\Sigma$, $j^\mu \neq 0$ in $V^-$,
$j^\mu = 0$ in $V^+$ and $j^\mu_-$ is tangential to $\Sigma$. (We
denote the limits to $\Sigma$ from inside by the index $-$.)
$j^\mu$ is a vector density; it is easy to show that $j^\mu$  is identically
conserved everywhere in $M$, $j^\mu_{\ ,\mu} = 0$.

The current $j^\mu(x)$ defines the spacetime four-velocity $u^\mu(x)$ and the
rest mole (particle) scalar density $n$ of the fluid in $V^-$ and at $\Sigma$
by
\be
  j^\mu = \sqrt{|g|}\,nu^\mu,
\label{1.2}
\ee
where $g := \det(g_{\mu\nu})$ and $g_{\mu\nu}u^\mu u^\nu = -1$. Hence,
\be
  n = \frac{1}{\sqrt{|g|}}\sqrt{-g_{\mu\nu}j^\mu j^\nu};
\label{1.3}
\ee
$n$ has a discontinuity of a step type at $\Sigma$.

In \cite{JK1}, it is shown that the fluid equations of motion
can be obtained from the Lagrange density $L_m$ which is given by
\be
  L_m = -\sqrt{|g|} n e(n),
\label{1.4}
\ee
where $e(n)$ is the energy per mole in the rest frame of the fluid and $L_m$
is considered as a function of $z^a$, $z^a_\mu$ and $g_{\mu\nu}$. As the
specific volume $V$ (i.e. volume of one mole in the rest frame) is
$1/n$, we obtain for the presure $p$ of the fluid
\be
  p = -\frac{\partial e}{\partial V} = n^2e'
\label{1.5}
\ee
in $V^-$.

\subsubsection{Stress energy density}
\label{sec:Tmunu}
By definition, the stress-energy tensor density of ideal fluid (see, eg.\
\cite{MTW}) has the form
\be
  T^{\mu\nu} = \sqrt{|g|}\left((\rho + p)u^\mu u^\nu + pg^{\mu\nu}\right),
\label{1.6}
\ee
where
\be
  \rho = ne(n)
\label{rho}
\ee
 is the rest mass density; $T^{\mu\nu}$ has a step
discontinuity at $\Sigma$. In this section, we collect some important formulas
valid for this tensor density.

Let us vary the action of the fluid
\be
  I_m = -\int_{V^-}d^4x\,\sqrt{|g|}ne(n)
\label{action-m}
\ee
with respect to $g_{\mu\nu}$. Using Eq.\ (\ref{1.3}), we have
\[
  \delta(\sqrt{|g|}n) = \delta\sqrt{-g_{\mu\nu}j^\mu j^\nu} =
  -\frac{1}{2\sqrt{|g|}}n^{-1}j^\mu j^\nu\delta g_{\mu\nu},
\]
which, together with the Eq.\ (\ref{1.2}) and the well-known variation formula
for determinats yields
\be
  \delta n = -\frac{1}{2}n(g^{\mu\nu} + u^\mu u^\nu)\delta g_{\mu\nu}.
\label{1.7}
\ee
Then,
\[
  \delta I_m = \int_{V^-}d^4x\sqrt{|g|} \left( -\frac{1}{2}g^{\mu\nu}ne +
  \frac{1}{2}n(g^{\mu\nu} + u^\mu u^\nu)(e + ne')\right)\delta g_{\mu\nu},
\]
and a straightforward calculation using Eqs.\ (\ref{rho}), (\ref{1.5}) and
(\ref{1.6}) leads to
\be
  T^{\mu\nu}(x) = 2\frac{\delta I_m}{\delta g_{\mu\nu}(x)}.
\label{1.8}
\ee

The next important relation is the Belinfante-Rosenfeld theorem
(\cite{bel1}, \cite{bel2} and \cite{ros}) applied to our
case: the Lagrange density $L_m$ must satisfy the following identity
\be
  \frac{\partial L_m}{\partial z^a_\mu}z^a_\nu + 2\frac{\partial
  L_m}{\partial g_{\mu\rho}}g_{\nu\rho} = L_m\delta^\mu_\nu.
\label{1.9}
\ee
This equation is equivalent to the requirement that $L_m$ is a
scalar density, and its derivation is straightforward. From the identity
(\ref{1.9}) and
the formula (\ref{1.8}), we obtain immediately that
\be
  T^\mu_\nu = L_m\delta^\mu_\nu - \frac{\partial L_m}{\partial
  z^a_\mu}z^a_\nu.
\label{1.10}
\ee
Thus, the so called canonical stress-energy tensor density
on the right hand side is equal to the source of gravitational field.

The formulas (\ref{1.8}) and (\ref{1.10}) imply the Noether identity:
\be
  \nabla_\mu T^\mu_\nu = \left( -\partial_\mu\frac{\partial L_m}{\partial
  z^a_\mu} + \frac{\partial L_m}{\partial z^a} \right)z^a_\nu.
\label{1.11}
\ee
There are counterparts to Eqs.\ (\ref{1.8})--(\ref{1.11}) within any
description of any type of ideal fluid. Derivation of Eq.\ (\ref{1.11}) starts
from the equation
\[
  \nabla_\mu T^\mu_\nu = \partial_\mu T^\mu_\nu -
  \Gamma^\rho_{\mu\nu}T^\mu_\rho.
\]
If one substitutes for $T^\mu_\nu$ from Eq.\ (\ref{1.10}) into the
first term on the right hand side and from Eq.\ (\ref{1.8}) into the second
one, the identity follows. One consequence of the Noether identity is that the
four components of the covector $\nabla_\mu T^\mu_\nu$ are not
independent:
\[
  j^\nu\nabla_\mu T^\mu_\nu = 0,
\]
because the definition of $j^\mu$ implies the identity $j^\mu z^a_\mu
\equiv 0$. Hence, the equation system $\nabla_\mu T^\mu_\nu = 0$ contains
only three independent equations (Euler equation); the energy
conservation equation,
\[
  \nabla_\mu(neu^\mu) = -p\nabla_\mu u^\mu,
\]
for the fluid is satisfied identically {\em within our description}.

\subsubsection{The variational formula}
\label{sec:variat}
Let us consider the
4-dimensional volume $V$ enclosed between two Cauchy surfaces $S_1$ and
$S_2$; the boundary $\Sigma$ of the fluid divides $V$ into two parts,
$V^-$ and $V^+$, and similarly $S_i$ into $S^-_i$ and $S^+_i$, $i =
1,2$. We assume that $S_1$ and $S_2$ are $C^1$-surfaces, that is the
induced metric on, as well as the unit normal vector to, $S_1$ and $S_2$
are both $C^1$.

Let us vary the matter action $I_m$ with respect to $z^a(x)$ and
$g_{\mu\nu}$; we obtain
\bea
  \delta I_m & = & \int_{\delta V^-}d^4x\, L_m + \int_{V^-}d^4x\, \partial_\mu
  \left(\frac{\partial L_m}{\partial z^a_\mu}\delta z^a \right) \nn \\
  & + & \int_{V^-}d^4x\,
  \left(\frac{\partial L_m}{\partial z^a} - \partial_\mu\frac{\partial
  L_m}{\partial z^a_\mu} \right) \delta z^a + \frac{1}{2} \int_{V^-}d^4x\,
  T^{\mu\nu}\delta g_{\mu\nu}.
\label{1.12}
\eea

The first two integrals can be transformed to surface integrals along $\Sigma$,
$S_1$ and $S_2$. For this aim, we use the coordinates $x^\mu$ in $M$ that are
adapted to the surfaces. This means that $x^0 = t_i$ along $S_i$, $i = 1,2$
and
\be
  F(z^a(x)) = x^3
\label{6.1}
\ee
along $\Sigma$, so that $x^k$, $k = 1,2,3$, are coordinates on $S_i$ and
$x^\alpha$, $\alpha = 0,1,2$ are coordinates on $\Sigma$. Then, the change of
$F$ if we vary $z^a$ is given by
\[
  F(z^a(x) + \delta z^a(x)) = x^3 + F_a\delta z^a,
\]
and the coordinate $x^3$ of $\Sigma$ changes by
\[
  \delta x^3 = - F_a\delta z^a|_{x^3 = 0}.
\]
Thus, we obtain for the first term:
\[
  \int_{\delta V^-}d^4x\, L_m = - \int_\Sigma dx^0dx^1dx^2\, L_m F_a\delta z^a.
\]
For the second term, we have
\[
  \int_{V^-}d^4x\, \partial_\mu \left(
  \frac{\partial L_m}{\partial z^a_\mu}\delta z^a \right) =
  \int_{\Sigma}d\Sigma\,
  \frac{\partial L_m}{\partial z^a_3}\delta z^a + \int_{S_2}dS\,
  \frac{\partial L_m}{\partial z^a_0}\delta z^a - \int_{S_1}dS\,
  \frac{\partial L_m}{\partial z^a_0}\delta z^a,
\]
where the abbreviations $d\Sigma = dx^0dx^1dx^2$ and $dS = dx^1dx^2dx^3$ are
used. Eq.\ (\ref{6.1}) implies that $F_az^a_\mu = \delta^3_\mu$, so we can
write
\[
  \frac{\partial L_m}{\partial z^a_3} = \frac{\partial L_m}{\partial z^a_\mu}
  F_bz^b_\mu.
\]
Collecting these results, we obtain the equation
\bea
  \delta I_m & = & \int_{V^-}d^4x\,
  \left(\frac{\partial L_m}{\partial z^a} - \partial_\mu\frac{\partial
  L_m}{\partial z^a_\mu} \right) \delta z^a
  + \int_\Sigma d\Sigma\,\left(\frac{\partial L_m}{\partial z^a_\mu}z^b_\mu -
  L_m\delta_a^b \right) F_b\delta z^a \nn \\
  & + & \int_{S_2}dS\,\frac{\partial L_m}{\partial z^a_0}\delta z^a
  - \int_{S_1}dS\,\frac{\partial L_m}{\partial z^a_0}\delta z^a
  + \frac{1}{2}\int_{V^-}d^4x\,T^{\mu\nu}\delta g_{\mu\nu}.
\label{7.3}
\eea
Thus, the field equations consist of {\em volume} equations that hold in $V^-$:
\be
  \frac{\partial L_m}{\partial z^a} - \partial_\mu\frac{\partial
  L_m}{\partial z^a_\mu} = 0,
\label{7.1}
\ee
and {\em surface} equations that hold at $\Sigma$:
\be
  \left(\frac{\partial L_m}{\partial z^a_\mu}z^b_\mu - L_m\delta^b_a
  \right) F_b = 0.
\label{7.2}
\ee
The surface of the star is an observer independent dynamical element of the
system.

Let us discuss the meaning of the field equations. For the volume equation
(\ref{7.1}), we just
invoke Noether's identity, Eq.\ (\ref{1.11}); we can then see that they are
equivalent to the conservation equations $\nabla_\mu T^\mu_\nu = 0$. The
surface equations can be rewritten as follows. First, using Eq.\
(\ref{1.10}), we have
\[
  T^\mu_\nu z^b_\mu = \left(-\frac{\partial L_m}{\partial
  z^a_\mu}z^b_\mu + L_m\delta^b_a\right)z^a_\nu.
\]
Hence,
\be
  T^\mu_\nu F_b z^b_\mu = \left(-\frac{\partial L_m}{\partial
  z^a_\mu}z^b_\mu + L_m\delta^b_a\right)F_b z^a_\nu.
\label{11.1}
\ee
However, $F_bz^b_\mu$ is covector normal to $\Sigma$, so the three
surface equations (\ref{7.2}) can be written in a covariant form as
\be
  T^\bot_\mu|_\Sigma = 0,
\label{12.1}
\ee
where $T^\bot_\mu = T^\nu_\mu \tilde{m}_\nu$ and $\tilde{m}_\nu$ is any
normal covector to $\Sigma$. Of these four equations, only three are
independent, because $T^\bot_\nu j^\nu$ is identically zero, as one easily
verifies; in fact, the three equations $T^\bot_k|_\Sigma = 0$ imply
(\ref{12.1}). It follows further from Eq.\ (\ref{1.6}) and from $F_bz^b_\mu
u^\mu = 0$ that Eq.\ (\ref{12.1}) is equivalent to the condition that the
pressure vanishes, $p = 0$, at the surface.

Formula (\ref{7.3}) is only valid in the adapted coordinates; in particular,
the Lagrangian density $L_m$ must be expressed in these coordinates. Let us
pass to more general coordinates. In fact, the
volume integrals are already in a covariant form, so we have just to transform
the boundary integrals. However, we will need explicitly only the integrals
over the Cauchy surfaces.

Let us define
\be
  p_a := \frac{\partial L_m}{\partial z^a_0}.
\label{momenta}
\ee
The reader can easily verify that Eq.\ (\ref{momenta}), which is written in
the adapted coordinates, defines a three-density $p_a$ along $S_i$
(independent of the adapted coordinates), because the quantity $L_m$ is a
four-density.

The field equations (both volume and surface) are, therefore, equivalent to
the formula
\[
  \delta \int_{V^-}d^4x\,L_m = \int_{S_2}d^3y\,p_a\delta z^a -
  \int_{S_1}d^3y\,p_a\delta z^a + \frac{1}{2}\int_{V^-}d^4x\,
  T^{\mu\nu}\delta g_{\mu\nu},
\]
and this formula is valid in any coordinates $x^\mu$ in $V^-$ and $y^k$, $k =
1,2,3$, along
$S_i$. Let us denote the matter occupied part of the Cauchy surface $x^0 = t$
by $S_t$ and the corresponding part of the matter space by $Z^-$. The
following coordinates will simplify all calculations: the intersection $S_t
\cap \Sigma$ is given by $x^3 = x^3(t)$ and $S_t$ by $x^3 < x^3(t)$; $x^k$, $k
= 1,2,3$, are coordinates on $S_t$; we will call them `time dependent adapted
coordinates.' Then, we can write
\[
  \delta \int_{V^-}d^4x\,L_m =
  \int_{t_1}^{t_2}dt\left(\frac{d}{dt}\int_{x^3<x^3(t)}d^3x\,p_a \delta z^a +
  \frac{1}{2}\int_{x^3<x^3(t)}d^3x\,T^{\mu\nu}\delta g_{\mu\nu}\right).
\]
If we define the Lagragian ${\cal L}_m$ by
\[
  {\cal L}_m = \int_{S_t}d^3x\,L_m
\]
and go to the limit $S_2 \rightarrow S_1$, we obtain the {\em variation
  formula} for the Lagrangian:
\be
  \delta {\cal L}_m = \int_{S_t}d^3x\,(p_a\delta z^a)^{\mbox{.}} +
  \int_{\partial S_t}d^2x\,p_a\delta z^a \dot{x}^3 +
  \frac{1}{2}\int_{S_t}d^3x\,T^{\mu\nu}\delta g_{\mu\nu}.
\label{varL}
\ee
The variation formula has been derived by careful inclusion of all `boundary
terms'; this will be the main strategy for our derivation of the Lagrangian
and Hamiltonian formalism that admit discontinuities. The role of the
variational formula is to generate all dynamical equations (including the
definition of momenta): the variation of
the L. H. S. is to be calculated and compared with the R. H. S.

Eq.\ (\ref{varL}) is also the point of departure for the transformation to
the Hamiltonian formalism.

\subsubsection{The Hamiltonian formalism}
\label{sec:legendre}
Let us disregard the surface term in the formula (\ref{varL}) and define the
Hamiltonian density $H_m$ by a Legendre transformation of the form
\[
  H_m := p_a \dot{z}^a - L_m.
\]
Then we obtain for the variation of the Hamiltonian ${\cal H}_m$, which is
defined by
\[
  {\cal H}_m := \int_{S_t}d^3x\,H_m,
\]
the relation
\[
 \delta {\cal H}_m = \delta\int_{S_t}d^3x\,p_a\dot{z}^a - \delta
 \int_{S_t}d^3x\,L_m.
\]
Performing carefully the variation in the first term and substituting from Eq.\
(\ref{varL}) for the second, we have
\bea
  \delta{\cal H}_m & = &
  \int_{S_t}d^3x\,(\delta p_a\dot{z}^a - \dot{p}_a \delta z^a)
  + \int_{x^3=x^3(t)}d^2x\,p_a(\dot{z}^a\delta x^3 - \dot{x}^3\delta z^a)
  \nn \\
  & - & \frac{1}{2}\int_{S_t}d^3x\,T^{\mu\nu}\delta g_{\mu\nu}.
\label{varH}
\eea
From this equation, not only the canonical equations are obtained, which
will consist of volume and of surface equations, but also the symplectic
structure of the system can be read off, which is given by the fist two
integrals on the R.H.S.; these integals can be interpreted as
$\omega(\delta z^a, \delta p_a; \dot{z}^a,\dot{p}_a)$, where $\omega$ is
the symplectic form and $(\delta z^a(x), \delta p_a(x))$,
$(\dot{z}^a(x),\dot{p}_a(x))$ are two vectors; notice that $\omega$ has
a surface part. This is a general and very important observation, which
will, for instance, help to decide what are the canonical variables for
the shell in the second part of the paper. The last integral in
(\ref{varH}) represents variation with respect to non-dynamical
`parameters' $g_{\mu\nu}(x)$.

We will need the following relations between the Hamiltonian density and the
stress-energy density:
\bea
  T^0_0 & = & -H_m,
\label{HT00} \\
  T^0_k & = & -p_a z^a_k,
\label{HT0k} \\
  T^k_l & = & \left(\frac{\partial H_m}{\partial p_a}p_a - H_m\right)
  \delta^k_l + \frac{\partial H_m}{\partial z^a_k}z^a_l,
\label{HTkl}
\eea
which are valid in the time dependent adapted coordinates. The first two
equations are obtained immediately from Eq.\ (\ref{1.10}) and the definition
of $p_a$. To derive the last equation, we first notice that Eq.\ (\ref{varH})
has the following consequence
\be
  T^{\mu\nu} = - 2\frac{\partial H_m}{\partial g_{\mu\nu}}.
\label{THg}
\ee
Second, we derive an equation analogous to (\ref{1.9}) for $H_m$; we use the
fact that $H_m$ behaves as a three-density if we change the coordinates $x^k$
keeping $t$ fixed and that $H_m$ is a function of $z^a$, $z^a_m$, $p_a$ and
$g_{\mu\nu}$, $H_m = H_m(z^a,z^a_\mu,p_a,g_{\mu\nu})$:
\[
  \frac{\partial H_m}{\partial z^a_l}z^a_k + \frac{\partial H_m}{\partial
  p_a}p_a\delta^l_k + 2\frac{\partial H_m}{\partial g_{\rho l}}g_{\rho k}
  = H_m\delta^l_k.
\]
Then, Eq.\ (\ref{HTkl}) follows immediately. Again, analogons of Eqs.\
(\ref{HT00})--(\ref{THg}) are valid for all types of ideal fluid.

To finish the Legendre transformation, we have also to express the velocity
$\dot{z}^a$ in terms of $p_a$, $z^a$ and $z^a_k$ in the Hamiltonian. This is
not completely straightforward. To begin with, we substitute for $L_m$ from
Eq.\ (\ref{1.4}) into the definition (\ref{momenta}) of $p_a$:
\[
  p_a = - \sqrt{|g|}\rho'\frac{\partial n}{\partial \dot{z}^a},
\]
where
\[
  \rho' = \frac{d\rho}{dn}.
\]
Eqs.\ (\ref{1.2}) and (\ref{1.3}) imply
\[
  \frac{\partial n}{\partial \dot{z}^a}
  = - \frac{1}{\sqrt{|g|}}u_\mu\frac{\partial j^\mu}{\partial\dot{z}^a}.
\]
The definition of $j^\mu$ implies the following identity (cf.\
\cite{JK1})
\be
 \frac{\partial j^\mu}{\partial z^a_\nu}z^a_\kappa = j^\mu\delta^\nu_\kappa -
  j^\nu\delta^\mu_\kappa.
\label{j-id}
\ee
Combining the three equations, we easily find
\be
  u_k = - \frac{p_az^a_k}{\rho'j^0},
\label{u}
\ee
where $j^0$ depends only on $z^a$ and $z^a_k$ (cf.\ (\ref{1.1})):
\[
  j^0 = h(z)\mbox{det}(z^a_k).
\]
The (3+1)-decomposition of the metric (see e.g.\ \cite{MTW}),
\[ \begin{array}{ll}
  g^{00} = - N^{-2},& g_{0k} = N_k, \\
  g_{kl} = q_{kl},& g = - qN^2,
\end{array} \]
gives, with the help of Eq. (\ref{1.2}),
\[
  q^{kl}u_ku_l = -1 - \frac{(u^0)^2}{g^{00}} = -1 +
  \frac{1}{N^2}\left(\frac{j^0}{\sqrt{q}}\right)^2.
\]
Substituting for $u_k$ from Eq. (\ref{u}), we obtain the identity
\be
  \frac{1}{n^2}\left(\frac{j^0}{\sqrt{q}}\right)^2 = 1 +
  \frac{q^{kl}z^a_kz^b_l}{\rho^{\prime 2}(j^0)^2}p_ap_b.
\label{n(p)}
\ee
This equation determines $n$ as a function of $p_a$, $z^a$ and $z^a_k$. The
solution depends on the unknown function $\rho(n)$ and is determined
only implicitly, in general.

The identity (\ref{j-id}) implies that
\[
  \frac{\partial j^k}{\partial \dot{z}^a} = - j^0x^k_a,
\]
where $x^k_a$ is the matrix inverse to $z^a_k$. As $j^k$ depends linearly on
$\dot{z}^a$, we have
\[
  j^k = - j^0x^k_a\dot{z}^a,
\]
or
\[
  \dot{z}^a = - \frac{z^k_aj^k}{j^0}.
\]
Substituting for $j^k$ from
\[
  j^k = g^{kl}(j_l - N_lj^0),
\]
for $j_l$ from (\ref{1.1}) and for $u_k$ from (\ref{u}), we obtain that
\[
  \dot{z}^a = N\frac{\sqrt{q}n}{\rho'(j^0)^2} q^{kl}z^a_kz^b_lp_b + N^kz^a_k.
\]
Then, Eqs.\ (\ref{1.10}) and (\ref{momenta}) yield
\be
  T^0_0 = -N\sqrt{q}\left(\rho +
  \frac{n}{\rho'(j^0)^2}q^{kl}z^a_kz^b_lp_ap_b\right) - N^kz^a_kp_a.
\label{valT00}
\ee
According to the formula (\ref{HT00}), this determines the form of the
Hamiltonian ${\mathcal H}_m$.

\subsubsection{The Euler picture}
\label{sec:invers}
At this stage, we have derived all important formulas of the Hamiltonian
formalism in which the fluid is described by the functions $z^a(x)$; these are
Lagrange coordinates (cf.\ \cite{brown}) and we can call the formalism `{\em
  Lagrange picture}'.  Sometimes the {\em Euler picture} is more practical,
however. This can be obtained by the following canonical transformation.  The
new fields $x^k(z,t)$ (Euler coordinates) are defined by
\be
x^k(z(x,t),t) =
x^k, \quad \forall x^k,t
\label{invers}
\ee
and the conjugate momenta $P_k$ by
\be
  P_k(z,t) := - X(z,t)z^a_k(x(z,t),t)p_a(x(z,t),t),
\label{Pp}
\ee
where
\[
  X := \mbox{det}\left(\frac{\partial x^k}{\partial z^a}\right).
\]
One easily checks that Eqs.\ (\ref{invers}) and (\ref{Pp}) define a canonical
transformation.

Let us first derive some useful relations. By differentiating Eq.
(\ref{invers}) with respect to $t$ at constant $x^k$, we obtain
\be
  \dot{x}^k = - x^k_a\dot{z}^a,
\label{dotx}
\ee
where
\[
  x^k_a := \frac{\partial x^k}{\partial z^a}.
\]
The derivative of the same equation with respect to $x^k$ at constant $t$
gives
\be
  x^k_az^a_l = \delta^k_l.
\label{gradx}
\ee
If the field $z^a(x,t)$ is changed to $\tilde{z}^a(x,t)$, then $x^k(z,t)$ is
changed to $\tilde{x}^k(z,t)$ satisfying
\[
  \tilde{x}^k(\tilde{z}(x,t),t) = x^k, \quad \forall x^k,t.
\]
Thus, if $\tilde{z}^a(x,t) = z^a(x,t) + \delta z^a(x,t)$, the above equation
implies that
\be
 \delta_*x^k = - x^k_a\delta z^a.
\label{deltax}
\ee
The symbol $\delta_*$ is to stress and to remind us that this variation is of
different kind than $\delta$, if applied to fields: the former is obtained by
comparing the values of the field at the same point of the matter space, that
is at different points of the spacetime; the latter compares the values of the
field at the same point of the spacetime.

With the help of the above relations, we can transform all formulas of the
Hamiltonian formalism.  Let us start with Eq.\ (\ref{varH}). First,
the inverse transformation for the momenta follows from Eq. (\ref{Pp}):
\be
  p_a(x,t) = - X^{-1}(x,t)x^k_a(z(x,t),t)P_k(z(x,t)t),
\label{inversP}
\ee
The time derivative of this equation at constant $x^k$ can be calculated with
the result
\bea
  \dot{p}_a & = & X^{-1}\left( x^k_ax^m_bz^b_{lm}P_k\dot{x}^l +
  z^b_lx^k_{ab}P_k\dot{x}^l + x^k_az^b_lP_k\partial_b\dot{x}^l
  + x^k_az^b_l\partial_bP_k\,\dot{x}^l \right. \nn \\
  & - & \left. P_l\partial_a\dot{x}^l - x^k_a\dot{P}_k \right),
\label{dotp}
\eea
where we introduced the abbreviation
\[
  z^a_{kl} := \frac{\partial^2 z^a}{\partial x^k \partial x^l}
\]
and similarly $x^k_{ab}$. Analogous formula holds for the variation $\delta
p_a$, one just have to replace dots by $\delta$'s. Employing these equations,
we obtain after a lenghty but straightforward calculation
\[
  \delta p_a\dot{z}^a - \dot{p}_a\delta z^a =
  X^{-1}(\delta_* P_k\dot{x}^k - \dot{P}_k\delta_*{x}^k)
  + X^{-1}\partial_a(P_kz^a_l\delta_*x^k\dot{x}^l)
  - X^{-1}\partial_a(P_lz^a_k\delta_*x^k\dot{x}^l),
\]
or
\bea
  \int_{S_t}d^3x\,(\delta p_a\dot{z}^a - \dot{p}_a\delta z^a)
  & = & \int_{Z^-}d^3z\,(\delta_*P_k\,\dot{x}^k - \dot{P}_k\delta_*x^k) \nn \\
  & + & \int_{Z^-}d^3z\,\partial_a\left((P_kz^a_l -
    P_lz^a_k)\delta_*x^k\dot{x}^l\right).
\label{inverseO}
\eea
Then, we transform the second integral on the R.H.S. of Eq.\ (\ref{varH}):
\beann
  \int_{x^3=x^3(t)}d^2x\,p_a(\dot{z}^a\delta x^3 - \dot{x}^3\delta z^a)
  & = & \int_{S_t}d^3x\,\partial_l\left(p_a(\dot{z}^a\delta_*x^l -
    \dot{x}^l\delta z^a)\right) \\
  & = & \int_{Z^-}d^3z\,Xz^b_l\partial_b\left(X^{-1}P_k(\dot{x}^k\delta_*x^l -
  \dot{x}^l\delta_*x^k)\right);
\eeann
we have used Eqs.\ (\ref{dotx}),(\ref{deltax}),(\ref{gradx}) and
(\ref{inversP}). Because of the identity $\partial_b(Xz^b_l) = 0$, we obtain
finally
\be
  \int_{x^3=x^3(t)}d^2x\,p_a(\dot{z}^a\delta x^3 - \dot{x}^3\delta z^a)
  = - \int_{Z^-}d^3z\,\partial_b\left((P_kz^b_l -
    P_lz^b_k)\delta_*x^k\dot{x}^l\right).
\label{surfaceO}
\ee
Eqs.\ (\ref{varH}), (\ref{inverseO}) and (\ref{surfaceO}) imply
\be
  \delta {\cal H}_m = \int_{Z^-}d^3z\,(\delta_*P_k\,\dot{x}^k -
  \dot{P}_k\delta_*x^k) + \frac{1}{2}\int_{Z^-}d^3z\,XT^{\mu\nu}\delta
  g_{\mu\nu}(x(x,t),t).
\label{varH'1}
\ee
Thus, the symplectic form has no surface term in the Euler picture. The
variation of the metric in the last term on the R.H.S. is independent of
the other variations, and it is defined by comparing values of the metric at
the same spacetime points.

Let us suppose that $\delta g_{\mu\nu}(x) = 0$, and let us introduce the
transformed Hamiltonian density $H'_m$ by $H'_m = XH_m$, so that
\[
  {\cal H}_m = \int_{Z^-}d^3z\,H'_m.
\]
Then,
\[
  \delta {\cal H}_m = \int_{Z^-}d^3z\,\delta_*H'_m,
\]
and we have
\be
  \int_{Z^-}d^3z\,\delta_*H'_m = \int_{Z^-}d^3z\,(\delta_*P_k\,\dot{x}^k -
  \dot{P}_k\delta_*x^k).
\label{varH'2}
\ee
In this form, the variational formula is suitable for derivation of the
canonical equations. To this aim, let us calculate $\delta_*H'_m$; $H'_m$ is of
the form $H'_m(x^k,x^k_a,P_k)$, hence
\[
  \delta_*H'_m = \left( \frac{\partial H'_m}{\partial x^k} - \partial_a
  \frac{\partial H'_m}{\partial x^k_a}\right)\delta_*x^k
  + \frac{\partial H'_m}{\partial P_k}\delta_*P_k
  + \partial_a\left(\frac{\partial H'_m}{\partial x^k_a}\delta_*x^k\right).
\]
Thus, the field equations consist of the volume equations:
\bea
  \dot{x}^k & = & \frac{\partial H'_m}{\partial P_k},
\label{field-x} \\
  - \dot{P}_k & = &  \frac{\partial H'_m}{\partial x^k} - \partial_a
  \frac{\partial H'_m}{\partial x^k_a},
\label{field-P}
\eea
and the surface equations:
\be
  F_a\frac{\partial H'_m}{\partial x^k_a}|_{\Sigma_z}
  = 0.
\label{field-surf}
\ee
Let us check that Eq.\ (\ref{field-surf}) is equivalent to (\ref{12.1}).  We
have
\[
  H'_m
= XH_m(z^a(x^l),z^a_k(x^l_b),p_a(z^a(x^l),z^a_k(x^l_b),P_k),g_{\mu\nu}(x^l)),
\]
so that
\[
  \frac{\partial H'_m}{\partial x^k_a}|_{x,P} = Xz^a_kH'_m
  + X\frac{\partial H'_m}{\partial z^b_l}|_{z,p}\frac{\partial z^b_l}{\partial
  x^k_a}
  + X\frac{\partial H'_m}{\partial p_b}|_{z,z_k}\frac{\partial p_b}{\partial
  x^k_a}.
\]
Eqs.\ (\ref{gradx}) and (\ref{inversP}) imply
\beann
  \frac{\partial z^b_l}{\partial x^k_a} & = & -z^b_kz^a_l, \\
  \frac{\partial p_b}{\partial x^k_a} & = & -p_bz^a_k + \delta^a_bp_cz^c_k,
\eeann
and we obtain easily
\[
  \frac{\partial H'_m}{\partial x^k_a}
  = Xz^a_l \left( \left(H_m
  - \frac{\partial H_m}{\partial p_b}p_b\right)\delta^l_k
  - \frac{\partial H_m}{\partial z^b_l}z^b_k\right)
  + X\frac{\partial H_m}{\partial p_a}p_bz^b_k.
\]
Application of Eqs.\ (\ref{HT0k}) and (\ref{HTkl}) as well as (\ref{varH})
simplify the expression to
\be
  \frac{\partial H'_m}{\partial x^k_a} = -Xz^a_\mu T^\mu_k.
\label{dHdx}
\ee
Hence, Eq.\ (\ref{field-surf}) becomes
\[
  F_az^a_\mu T^\mu_k|_{\Sigma_z} = 0,
\]
which {\em is} equivalent to (\ref{12.1}).

Finally, the transformed Eq.\ (\ref{n(p)}) reads
\be
  n^2 + \frac{n^2}{h^2\rho^{\prime 2}}q^{kl}P_kP_l = \frac{h^2}{qX^2}.
\label{n(P)}
\ee
For example, in the case of dust, $\rho = \mu n$, where $\mu$ is a
constant (rest mass per mole or particle) and Eq.\ (\ref{n(P)}) can be solved
explicitly:
\[
  n = \frac{1}{\sqrt{q}X} \frac{\mu h^2}{\sqrt{\mu^2h^2 + q^{kl}P_kP_l}}.
\]
For dust, Eq.\ (\ref{valT00}), which determines the form of the matter
Hamiltonian, specializes to
\be
  T^0_0 = -N\sqrt{\mu^2(j^0)^2 + q^{kl}z^a_kz^b_lp_ap_b} - N^kz^a_kp_a.
\label{T00dust}
\ee

\subsection{The gravity becomes dynamical}
In the foregoing sections, gravity was just an external field. Here, it will
become dynamical: the metric $g_{\mu\nu}(x)$ will satisfy Einstein's equations
with the fluid stress energy tensor as a source.

\subsubsection{Description of the system}
\label{sec:descr-step}
The main problem which we shall meet is the following. If Einstein's equations
are satisfied, the discontinuity in the distribution of the fluid leads to a
discontinuity in derivatives of the metric. Thus, we must allow for such
discontinuity from the very beginning. Moreover, a general variation of the
metric, which includes a shift of the coordinates of the discontinuity, will
have a jump of higher order than the metric itself: if the second derivatives
of the metric have a jump, then the first derivative of its variation will
have a jump, etc. If we write naively the usual expression for the variation
of the action in the case of delta-function fluid distribution, then many
terms in it look meaningless within the theory of distributions
($\delta$-functions multiplied by discontinuous functions, etc.). 
Some ingenious calculation of all variations might still lead to meaningful
expressions. Instead, we resort to a simple trick by which the problem is
avoided: we fix the spacetime
coordinates of the discontinuity surface $\Sigma$. In this way, the surface of
the discontinuity is formally made to an `externally given' boundary. The
fields $z^a(x)$ and $g_{\mu\nu}(x)$ will satisfy simple boundary conditions at
$\Sigma$, and these conditions will be `inherited' by their variations. Such a
strategy is possible within the general relativity, because it can be
considered as a partial fixing of gauge. Indeed, any change of the coordinates
of the discontinuity surface $\Sigma$ can be considered as a superposition of
a transformation of coordinates in a neighbourhood of $\Sigma$ keeping the
physical fields fixed, and a change of the physical fields keeping the
coordinates fixed; the first step is just a change of gauge. The dynamics of
the surface is determined by the form of the metric near and at the surface.

To be more specific about the boundary conditions, let us choose the
coordinates $z^a$ in $Z$ such that $\Sigma_z$ is given by $z^3 = 0$, and the
coordinates $x^\mu$ in $M$ such that $\Sigma$ is defined by $x^3 = 0$. Thus,
for the matter fields, we require
\be
 z^3|_\Sigma = 0, \quad \delta z^3|_\Sigma = 0.
\label{boundcond1}
\ee
It follows that
\be
 x^3|_{\Sigma_z} = 0, \quad \delta x^3|_{\Sigma_z} = 0
\label{boundcond2}
\ee
in the Euler picture. We further assume that
\begin{description}
\item[Condition 1] the spacetime $(M,g)$ is asymptotically flat and globally
  hyperbolic;
\item[Condition 2] the metric $g_{\mu\nu}(x)$ is piecewise $C^\infty$ in $M$,
  the only discontinuity being that its second derivatives jump at
  $\Sigma$.
\end{description}
Then the variation $\delta g_{\mu\nu}(x)$ satisfies analogous Condition 2.

The total action for our fluid-gravity system is $I = \bar{I}_m + I_g$. Here
\[
  I_g = \frac{1}{16\pi\mbox{G}}\int_Vd^4x\,\sqrt{|g|}R,
\]
where G is the Newton constant and $R$ is the curvature scalar of
$g_{\mu\nu}$. The function $R(x)$ can have a step discontinuity at $\Sigma$.
$\bar{I}_m$ is obtained from $I_m$
of Eq. (\ref{action-m}) after the following substitution
\be
  z^3(x^\alpha,x^3=0) = 0, \quad
  \dot{z}^3(x^\alpha,x^3=0) = 0, \quad
  z^3_A(x^\alpha,x^3=0) = 0,
\label{substitut}
\ee
$\alpha = 0,1,2$ and $A = 1,2$. Thus, $\bar{I}_m$ contains less variables than
$I_m$.

The integration volume $V$ is chosen to be bounded by two Cauchy surfaces,
$S_1$ and $S_2$, and by a timelike surface $\Sigma^+$ (which will be
eventually pushed to the infinity). Let the coordinates $x^\mu$ be adapted
also to $\Sigma^+$ so that $\Sigma^+$ is defined by $x^3 = r^+$. The matter
boundary $\Sigma$ divides $V$ into $V^-$ and $V^+$, and $S_i$ into $S_i^-$ and
$S_i^+$.

\subsubsection{The variational formula}
The variation of the gravity action $I_g$ can be obtained from the following
fundamental lemma that has been shown in \cite{JK2}.
\begin{lem}
Let the integration volume $V$ of the action $I_g$ be bounded by two spacelike
surfaces $S_1$ and $S_2$, and by a smooth timelike surface $\Sigma$; let
$x^\mu$ be some coordinates in $V$, $y^k$ in $S_i$, $\xi^\alpha$ in $\Sigma$
and $\eta^A$ in $\partial S_i = \Sigma\cap S_i$. Then
\bea
  \delta I_g
 & = & - \frac{1}{16\pi\mbox{G}}\int_Vd^4x\,G^{\mu\nu}\delta g_{\mu\nu} \nn \\
 & - & \frac{1}{16\pi\mbox{G}}\int_{S_2}d^3y\,q_{kl}\delta\pi^{kl}
 + \frac{1}{16\pi\mbox{G}}\int_{S_1}d^3y\,q_{kl}\delta\pi^{kl} \nn \\
 & + & \frac{1}{8\pi\mbox{G}}\int_{\partial S_2}d^2\eta\,
 \sqrt{\lambda}\delta\alpha - \frac{1}{8\pi\mbox{G}}\int_{\partial S_1}
 d^2\eta\,\sqrt{\lambda}\delta\alpha \nn \\
 & - & \frac{1}{16\pi\mbox{G}}\int_\Sigma d^3\xi\,\gamma_{\alpha\beta}\delta
 Q^{\alpha\beta},
\label{varIg}
\eea
where
\be
  G^{\mu\nu} := \sqrt{|g|}(R^{\mu\nu} - \frac{1}{2}g^{\mu\nu}R),
\label{GR}
\ee
$R^{\mu\nu}$ is the Ricci tensor of the metric $g_{\mu\nu}$, $q_{kl}$ is the
induced metric on $S_i$ written with respect to the coordinates $y^k$, $q$ its
determinant,
\beann
  \pi^{kl} & := & \sqrt{q}(Kq^{kl} - K^{kl}), \\
  K_{kl} & = & - n_{\mu;\nu} \frac{\partial x^\mu}{\partial y^k}\frac{\partial
  x^\nu}{\partial y^l}, \\
  K & = & q^{kl}K_{kl},
\eeann
$n_\mu$ is the future directed unit normal to $S_i$ so that $K_{kl}$ is the
second fundamental form of the surface $S_i$, $\gamma_{\alpha\beta}$ is the
metric induced on $\Sigma$ written with respect to the coordinates
$\xi^\alpha$, $\gamma$ its determinant
\bea
  Q^{\alpha\beta} & := & \sqrt{|
  \gamma|}(L\gamma^{\alpha\beta} - L^{\alpha\beta}),
\label{QL} \\
  L_{\alpha\beta} & = & \tilde{m}_{\mu;\nu }\frac{\partial x^\mu}{\partial
  \xi^\alpha}\frac{\partial x^\mu}{\partial \xi^\beta}, \nn \\
  L & = & \gamma^{\alpha\beta}L_{\alpha\beta}, \nn
\eea
$\tilde{m}_\mu$ is the external (with respect to the volume $V$) unit normal to
$\Sigma$ so that $L_{\alpha\beta}$ is the second fundamental form of $\Sigma$,
$\lambda$ is the determinant of the 2-metric $\lambda_{AB}$ induced on
$\partial S_i$ written with respect to the coordinates $\eta^A$ and $\alpha$
  is defined by
\[
  \alpha := - \mbox{\em arcsinh}(g_{\mu\nu}n^\mu\tilde{m}^\nu).
\]
\end{lem}

The Lemma 1 is completely general, independent of the form and description of
the matter; it determines the `gravitational part' of the variation formula
that we are going to derive. For the `matter part', we can use the formula
(\ref{7.3}) in which the surface integral along $\Sigma$ is left out. Indeed,
$F(z) = z^3$ for our special coordinates and the boundary condition
(\ref{boundcond1}) gives $F_a\delta z^a = 0$. Hence,
\bea \delta I_m & = &
\int_{V^-}d^4x\, \left(\frac{\partial\bar{L}_m}{\partial z^a} -
  \partial_\mu\frac{\partial \bar{L}_m}{\partial z^a_\mu} \right) \delta z^a +
\int_{S_2^-}dS\,\frac{\partial\bar{L}_m}{\partial z^a_0}\delta z^a
- \int_{S_1^-}dS\,\frac{\partial\bar{L}_m}{\partial z^a_0}\delta z^a \nn \\
& + & \frac{1}{2}\int_{V^-}d^4x\,T^{\mu\nu}\delta g_{\mu\nu}.
\label{varIm}
\eea

Our next task is to rewrite the surface integrals in Eq. (\ref{varIm}) in a
covariant way. We define, in analogy with Eq.\ (\ref{momenta}),
\[
  \bar{p}_a := \frac{\partial\bar{L}_m}{\partial\dot{z}^a}.
\]
By a similar argument as in Sec. \ref{sec:variat}, $\bar{p}_a$ are surface
densities, and the covariant forn of the integrals is:
\be
  \int_{S_2^-}dS\,\frac{\partial\bar{L}_m}{\partial z^a_0}\delta z^a
  - \int_{S_1^-}dS\,\frac{\partial\bar{L}_m}{\partial z^a_0}\delta z^a
  = \int_{S_2^-}d^3y\,\bar{p}_a\delta z^a
  - \int_{S_1^-}d^3y\,\bar{p}_a\delta z^a.
\label{surfm}
\ee

The relations between the old and new matter momenta will play some role. They
can be summarized as follows: in $V^-$, we simply have
\be
  \bar{p}_a = p_a,
\label{pinV}
\ee
whereas at $\Sigma$,
\be
  \bar{p}_A = p_A, \quad  p_3 = p(\bar{p}_A, z^a, z^a_k),
\label{patS}
\ee
where $p$ is some function of the variables indicated.
Eq.\ (\ref{pinV}) and the first Eq.\ of (\ref{patS}) follow directly from the
definitions, if the substitutions (\ref{substitut}) is made in the expressions
on the R.H.S.'s. As $\bar{L}_m$ does not depend on $\dot{z}^3$ at the
boundary, there is {\em no} $\bar{p}_3|_\Sigma$; $p_3|_\Sigma$ as given by
Eq.\ (\ref{momenta}) with the substitutions (\ref{substitut}) is, however,
non-zero and it can be expressed as in the second Eq.\ of (\ref{patS}). Let us
give a proof.  The solution of Eq.\ (\ref{momenta}) with respect to
$\dot{z}^a$ reads
\[
  \dot{z}^a = \dot{z}^a(p_a, z^a, z^a_k).
\]
At $\Sigma$, we must have $\dot{z}^3 = 0$, so we obtain one constraint for the
functions $p_a|_\Sigma$:
\be
  \dot{z}^3(p_a, z^a, z^a_k) = 0.
\label{con-p}
\ee
This can be solved for $p_3|_\Sigma$; the second Eq.\ of (\ref{patS}) is the
solution.

As an example, we work out the explicit form of Eq.\ (\ref{con-p}) for the
dust. We easily obtain from Eq.\ (\ref{1.1})
\[
  j^k = -hX^{-1}x^k_a\dot{z}^a,
\]
if we observe that the intermediately resulting terms can be expressed by
means of $z^a_k$-derivatives of the determinant of the matrix $z^a_k$. Eq.\
(\ref{1.3}) yields
\[
  n = \frac{h}{X\sqrt{|g|}}\sqrt{-g_{00} + 2\tilde{N}_{a}\dot{z}^a -
  \tilde{q}_{ab}\dot{z}^a\dot{z}^b},
\]
where
\beann
  \tilde{N}_a & := & N_kx^k_a, \\
  \tilde{q}_{ab} & := & g_{kl}x^k_ax^l_b.
\eeann
Then, for $\rho = \mu n$, we have from (\ref{momenta}) and (\ref{1.4}):
\[
  p_a = - \mu\frac{h^2}{X^2\sqrt{|g|}}\frac{1}{n}(\tilde{N}_{a} -
  \tilde{q}_{ab}\dot{z}^b),
\]
so that
\[
  \dot{z}^a = \tilde{q}^{ab}\left(\tilde{N}_b + \frac{X^2\sqrt{|g|}}{\mu
  h^2}np_b\right),
\]
and the desired constraint (\ref{con-p}) reads
\be
  \tilde{N}^3 + n\frac{X^2N\sqrt{q}}{\mu h^2}\tilde{q}^{3a}p_a = 0,
\label{constraint}
\ee
where
\[
  \tilde{q}^{ab} := q^{kl}z^a_kz^b_l
\]
and
\[
  \tilde{N}^a := N^kz^a_k.
\]
For $n$, we have to insert from Eq.\ (\ref{n(p)}):
\[
  \frac{1}{n} = \frac{X^2\sqrt{q}}{\mu h^2}\sqrt{\mu^2h^2X^{-2} +
  \tilde{q}^{ab}p_ap_b}.
\]
Thus, the constraint (\ref{constraint}) can be written as a quadratic equation
for $p_3$, whose general
solution is
\[
  p_{3\pm} = - \frac{\tilde{q}^{3A}}{\tilde{q}^{33}}p_A \pm \tilde{N}^3
  \sqrt{\frac{\tilde{\lambda}^{AB}p_Ap_B +
  \mu^2h^2X^{-2}}{\tilde{q}^{33}\left(N^2\tilde{q}^{33} -
  (\tilde{N}^3)^2\right)}},
\]
where
\[
  \tilde{\lambda}^{AB} = \lambda^{CD}z^A_Cz^B_D
\]
and
\[
  \lambda_{AB} = q_{AB}
\]
is the metric induced on $\partial S$. Only the lower sign is admissible, as we
can easily see from Eq.\ (\ref{constraint}). Hence, finally,
\[
  p_3|_\Sigma = - \frac{\tilde{q}^{3A}}{\tilde{q}^{33}}p_A - \tilde{N}^3
  \sqrt{\frac{\tilde{\lambda}^{AB}p_Ap_B +
  \mu^2h^2X^{-2}}{\tilde{q}^{33}\left(N^2\tilde{q}^{33} -
  (\tilde{N}^3)^2\right)}}.
\]

Eqs.\ (\ref{varIg}), (\ref{varIm}) and (\ref{surfm}) imply the final
{\em variational formula} for our gravity-fluid system:
\bea
  \delta I & = & \int_{V^-}d^4x\,
  \left(\frac{\partial\bar{L}_m}{\partial z^a} - \partial_\mu\frac{\partial
  \bar{L}_m}{\partial z^a_\mu} \right) \delta z^a
  - \frac{1}{16\pi\mbox{G}}\int_Vd^4x\,G^{\mu\nu}\delta g_{\mu\nu} \nn \\
  & + & \int_{S_2^-}d^3y\,\bar{p}_a\delta z^a
  - \int_{S_1^-}d^3y\,\bar{p}_a\delta z^a \nn \\
  & - & \frac{1}{16\pi\mbox{G}}\int_{S_2}d^3y\,q_{kl}\delta\pi^{kl}
  + \frac{1}{16\pi\mbox{G}}\int_{S_1}d^3y\,q_{kl}\delta\pi^{kl} \nn \\
  & + & \frac{1}{8\pi\mbox{G}}\int_{\partial
  S_2}d^2\eta\,\sqrt{\lambda}\delta\alpha
  - \frac{1}{8\pi\mbox{G}}\int_{\partial
  S_1}d^2\eta\,\sqrt{\lambda}\delta\alpha \nn \\
 & - & \frac{1}{16\pi\mbox{G}}\int_\Sigma d^3\xi\,\gamma_{\alpha\beta}\delta
 Q^{\alpha\beta} + \frac{1}{2}\int_{V^-}d^4x\,T^{\mu\nu}\delta g_{\mu\nu}.
\label{varI}
\eea

From the formula (\ref{varI}) we can read off the field equations;
within $V^-$, we have:
\bea
  \frac{\partial L_m}{\partial z^a} - \partial_\mu\frac{\partial
  L_m}{\partial z^a_\mu} & = & 0,
\label{matter-eq} \\
  G^{\mu\nu} & = & 8\pi\mbox{G}T^{\mu\nu},
\label{einstein-}
\eea
and within $V^+$, we have:
\be
  G^{\mu\nu} = 0.
\label{einstein+}
\ee
Apparently, the surface equation (\ref{7.2}) has been lost. However, using
the boundary Condition 2, we easily find that $G^\bot_\mu$ is continuous at
$\Sigma$.  Hence,
\[
  \lim_{x^3=0-}G^\bot_\mu = 0
\]
and the surface field equation follows from Eq. (\ref{einstein-}). We also
observe that the dynamics can be completely shifted to the gravity if
the ideal fluid is described by co-moving coordinates everywhere in $V^-$.

Putting everything together, we obtain in an analogous way as in Sec.
\ref{sec:variat}: the field equations are equivalent to the following relation
(which is an analogon of Eq.\ (5.16) of \cite{JK2})
\bea
  \delta{\cal L} & = & \int_{S^-}d^3y\,(\bar{p}_a\delta z^a)^{\mbox{.}} -
  \frac{1}{16\pi\mbox{G}}\int_{S}d^3y\, (q_{kl}\delta\pi^{kl})^{\mbox{.}}
  + \frac{1}{8\pi\mbox{G}}\int_{\partial S}d^2\eta\,
  (\sqrt{\lambda}\delta\alpha)^{\mbox{.}} \nn \\
  & - & \frac{1}{16\pi\mbox{G}}\int_{\partial S}d^2\eta\,
  \gamma_{\alpha\beta}\delta Q^{\alpha\beta},
\label{varLtot}
\eea
where ${\cal L}$ is the Lagrangian of the system,
\[
  {\cal L} := \int_SdS\,\bar{L}.
\]
Eq.\ (\ref{varLtot}) is the {\em variation formula} for our system.

\subsubsection{The Hamiltonian formalism}
\label{sec:legendr-gstep}
Eq.\ (\ref{varLtot}) is a good starting point for the Legendre
transformation to a Hamiltonian formalism. We define one of the conceivable
total Hamiltonians for our system by
\be
  \check{\cal H} := \int_{S^-}d^3y\,\bar{p}_a\dot{z}^a
  - \frac{1}{16\pi\mbox{G}}\int_Sd^3y\,q_{kl}\dot{\pi}^{kl}
  + \frac{1}{8\pi\mbox{G}} \int_{\partial S}d^2\eta\,\sqrt{\lambda}\dot{\alpha}
  - {\cal L}.
\label{defH}
\ee
Then, the field equations can be obtained from the {\em variation formula}
\bea
  \delta\check{\cal H}
  & = & \int_{S^-}d^3y\,(\dot{z}^a\delta\bar{p}_a - \dot{\bar{p}}_a\delta z^a)
  + \frac{1}{16\pi\mbox{G}}\int_{S}d^3y\,(\dot{q}_{kl}\delta\pi^{kl}
  - \dot{\pi}^{kl}\delta q_{kl}) \nn \\
  & + & \frac{1}{8\pi\mbox{G}}\int_{\partial S}d^2\eta\,
  \frac{1}{2\sqrt{\lambda}}(\dot{\alpha}\delta\lambda - \dot{\lambda}\delta
  \alpha) + \frac{1}{16\pi\mbox{G}}\int_{\partial S}d^2\eta\,
  \gamma_{\alpha\beta}\delta Q^{\alpha\beta}.
\label{varHtot}
\eea

To find the explicit form of the Hamiltonian we use the lemma
\begin{lem}
In the adapted coordinates $x^\mu$ defined in Sec.\ \ref{sec:descr-step}, the
following identity holds at any $S = S_t$ in the volume $V$
\bea
  \lefteqn{\int_SdS\,q_{kl}\dot{\pi}^{kl}
  - 2\int_{\partial S}d\partial S\,\sqrt{\lambda}\dot{\alpha}= } \nn \\
  & - & 2\int_SdS\,\sqrt{|g|}R^0_0 - 2\int_{\partial S}d\partial
  S\,\sqrt{|\gamma|}L^0_0,
\label{ident}
\eea
where $L^{\alpha\beta}$ is the second fundamental form of the boundary
$\partial V$ corresponding to the normal oriented outwards from $V$, $dS =
dx^1dx^2dx^3$ and $d\partial S = dx^1dx^2$.
\end{lem}
The derivation of this identity is given in \cite{JK2} (Eq.\ (6.3)); the form
(\ref{ident}) is easily obtained if one uses the equation $Q^{AB}g_{AB} -
Q^{00}g_{00} = \sqrt{|\gamma|}L^0_0$, which follows from Eq.\ (\ref{QL}). We
also observe that
\[
  \int_{S^-}dS\,\bar{p}_a\dot{z}^a - \bar{\cal L}_m = \bar{\cal H}_m,
\]
where $\bar{\cal H}_m$ is obtained from ${\cal H}_m$ as given in the Sec.\
\ref{sec:legendre} by the substitutions (\ref{substitut}), (\ref{pinV}) and
(\ref{patS}). Thus, the substitution for $p_3$ is discontinuous near $\Sigma$.
It could, therefore, seem that the corresponding Hamiltonian density
$\bar{H}_m$ would not be continuous at $\Sigma$, but this is not true.
The reason is that $H_m|_\Sigma$ does not depend on $p_3$, if the conditions
(\ref{boundcond1}) are satisfied:
\[
  \left.\frac{\partial H_m}{\partial p_a}\right|_\Sigma = \dot{z}^3|_\Sigma =
  0.
\]
Hence, we have from Eq.\ (\ref{HT00}):
\[
  \bar{\cal H}_m = - \int_SdS\,T^0_0.
\]
Collecting all results, we obtain finally:
\be
  \check{\cal H} = \int_SdS\,\left(- T^0_0 + \frac{1}{8\pi\mbox{G}}G^0_0\right)
  + \frac{1}{8\pi\mbox{G}}\int_{\partial S}d\partial S\,\sqrt{|\gamma|}L^0_0.
\label{formHtot}
\ee
This is the full `off-shell' Hamiltonian of our system. If Einstein's
equations hold, its value is just the surface integral.

Let us rewrite the volume integral in a covariant form. Any tensor density
$W^\mu_\nu$ satisfies the identity
\[
  W^0_0 = \frac{1}{g^{00}}(W^{00} - g^{0k}W^0_k).
\]
The unit future-oriented normal covector $n_\mu$ to $S$ has the components
$n_\mu = - N\delta^0_\mu$ with respect to the adapted coordinates; it follows
that
\be
  W^0_0 = - \sqrt{q}(Nw^{\bot\bot} + N^kw^\bot_k),
\label{we00}
\ee
where
\beann
  w^{\bot\bot} & := & \frac{1}{\sqrt{|g|}} W^{\mu\nu}n_\mu n_\nu, \\
  w^\bot_k & := & \frac{1}{\sqrt{|g|}} W^\mu_kn_\mu.
\eeann
Thus, the volume integral can be written as
\be
  \int_SdS\,\left(-T^0_0 + \frac{1}{8\pi\mbox{G}}G^0_0\right)
  = \frac{1}{8\pi\mbox{G}}\int_Sd^3y\,\sqrt{q}(NC + N^kC_k),
\label{NC}
\ee
where
\be
  C := -\frac{1}{\sqrt{|g|}} (G^{\mu\nu} - 8\pi\mbox{G}\,T^{\mu\nu})n_\mu n_\nu
\label{superham}
\ee
and
\be
  C_k := - \frac{1}{\sqrt{|g|}} (G^\mu_k - 8\pi\mbox{G}\,T^\mu_k)n_\mu
\label{supermom}
\ee
are the {\em super-Hamiltonian} and the {\em supermomentum} of our system (or
scalar and vector constraint functions).

Let us return to the formula (\ref{varHtot}), which not only implies the field
equations, if we perform the variation on the L.H.S. and compare the result
with the R.H.S., but it also determines the so-called {\em control mode} (see,
eg.\ \cite{K-T}) and the type of boundary value problem for the field
equations. We observe that this mode is a kind of ``curvature-control-mode'';
it amounts to keeping fixed (controling) the external curvature
$Q^{\alpha\beta}$ at the boundary $\Sigma^+$ (see \cite{JK2}). Such a boundary
problem for Einstein's equations has not been studied. To pass to a more
natural and in fact conventional approach, we have to perform an additional
Legendre transformation at the boundary \cite{JK2}:
\[
  {\cal H} = \check{\cal H}
  - \frac{1}{16\pi\mbox{G}}\int_{\partial S}d^2\eta\,\gamma_{AB}Q^{AB}
\]
so that we have, using also Eq.\ (\ref{NC}), finally
\be
  {\cal H} = \frac{1}{8\pi\mbox{G}}\int_SdS\,\sqrt{q}(NC + N^kC_k)
  - \frac{1}{16\pi\mbox{G}}\int_{\partial S}d\partial S\,Q^{00}\gamma_{00}
\label{formHfin}
\ee
and
\bea
  \delta{\cal H}
  & = & \int_{S^-}dS\,(\dot{z}^a\delta\bar{p}_a - \dot{\bar{p}}_a\delta z^a)
  + \frac{1}{16\pi\mbox{G}}\int_{S}dS\,(\dot{q}_{kl}\delta\pi^{kl}
  - \dot{\pi}^{kl}\delta q_{kl}) \nn \\
  & + & \frac{1}{8\pi\mbox{G}}\int_{\partial S}d\partial S\,
  (\dot{\alpha}\delta\lambda - \dot{\lambda}\delta \alpha) \nn \\
  & + & \frac{1}{16\pi\mbox{G}}\int_{\partial S}d\partial S\,
  (\gamma_{00}\delta Q^{00} + 2\gamma_{0A}\delta Q^{0A}
  - Q^{AB}\delta\gamma_{AB}).
\label{varHfin}
\eea
The last surface integral in Eq.\ (\ref{formHfin}) will result in the
A.D.M. energy, if the limit $\Sigma^+ \rightarrow \infty$ is carefully
performed (this has been shown in \cite{JK2}). The last one in Eq.\
(\ref{varHfin}) defines the way of controle: $Q^{00}$, $Q^{0A}$ and
$\gamma_{AB}$ are kept fixed at the boundary.

The transformation to the Euler picture in the matter part of the
Hamiltonian is straightforward; let us denote the resulting Hamiltonian
density by $\bar{H}'_m$. Most formulas of Sec.\ \ref{sec:invers} will result
in the analogous formulas for $\bar{H}'_m$, if the substitutions
(\ref{substitut}), (\ref{pinV}) and  (\ref{patS}) are performed in them. For
example, we have to use the modified formula (\ref{HT0k}), which will read at
$\Sigma$:
\[
  T^0_A = - \bar{p}_Bz^B_A, \quad T^0_3 = - \bar{p}_Bz^B_3 -
  p(\bar{p},z)z^3_3,
\]
etc. Only those formulas that contain derivatives of $H'_m$ with respect to
the variables which are not contained in $\bar{H}'_m$ (like $p_3|_\Sigma$ and
$x^3_A|_\Sigma$) need some care to be properly transformed.

We observe finally that the field equations derived from Eq.\ (\ref{varHfin})
(or an analogous equation of the Euler picture) will have the form of {\em
canonical equations}. This is interesting, because Eq. (\ref{field-surf}) does
not seem to have such form. The variations in Eq.\ (\ref{varHfin}) (or that of
the Euler picture) must satisfy the boundary conditions (\ref{boundcond1})
(or (\ref{boundcond2})) and will, therefore, lead to trivial surface
equations; the volume equations alone {\em have} the canonival form.  For
example, the would be counterpart of Eq.\ (\ref{field-surf}) originates from
the term
\[
  \left.\frac{\partial \bar{H}'_m}{\partial x^k_3}\delta x^k\right|_\Sigma
\]
in the variation of $\bar{H}'_m$. As $\delta x^3|_\Sigma = 0$, the only
equation which is implied thereby reads
\[
  \left.\frac{\partial \bar{H}'_m}{\partial x^A_3}\right|_\Sigma = 0.
\]
Further, a counterpart of Eq.\ (\ref{dHdx}) is valid for this derivative, so
the above equation is equivalent to
\[
  z^3_\mu T^\mu_A|_\Sigma = 0.
\]
Moreover, $z^3_B|_\Sigma = 0$, so the equation reduces to $T^3_A|_\Sigma = 0$.
However, these components of $T^\mu_\nu$ vanish identically at $\Sigma$,
besause $u^3|_\Sigma = 0$ (cf. Eq.\ (\ref{1.6})). Thus, there is no surface
field equation.

\section{Fluid shell}
In this section, we are going to describe the dynamics of a delta-function
distribution of fluid. The matter will be coupled to the dynamical gravity
from the start. We shall consider a special case: just one shell in
vacuum; a generalization to more shells surrounded by a piecewise smooth
matter is straightforward as far as the shells do not intersect.

\subsection{Action in the volume form}
The shell can be represented as a delta-function singularity in the mole
density $h$. The action can then be written as a volume integral of the same
form as for a regular distribution of matter. This holds also for the
gravitational part. We shall give a more detailed description of this volume
form and then transform it to a combinations of volume and surface integrals,
where no delta-functions will feature. This may be useful, because much more
general choice of coordinates is then allowed. Indeed, the $\delta$-function
method works only if the coordinates are such that the corresponding
components of the four-metric are continuous. Further, the coordinate position
of the shell---the three-surface in the spacetime $M$ and the two-surface in
the matter space $Z$---will be kept fixed. Here, everything can be repeated
what has already been said in Sec.\ \ref{sec:descr-step} about this point.

The matter space $Z$ remains, therefore, three-dimensional first. Let
$\Sigma_z$ be a two-dimensional surface in $Z$ on which the matter is
concentrated. Let $z^a$ be coordinates adapted to $\Sigma_z$ so that the
equation $z^3 = z^3_0$ determines $\Sigma_z$. Such coordinates are determined
up to a transformation
\[
\begin{array}{ll}
  z^{\prime 1} = z^{\prime 1}(z^a), & z^{\prime 2} = z^{\prime 2}(z^a), \\
  z^{\prime 3} = z^{\prime 3}(z^a), & z^{\prime 3}_0 = z^{\prime 3}(z^A,z^3_0),
\end{array}
\]
where $z^{\prime 3}_0$ is a constant independent of $z^A$.
Then,
\[
  \frac{\partial(z^1,z^2,z^3)}{\partial(z^{\prime 1},z^{\prime 2},z^{\prime
  3})} = \frac{\partial(z^1,z^2)}{\partial(z^{\prime 1},z^{\prime 2})}
  \cdot \frac{\partial z^3}{\partial z^{\prime 3}}.
\]
We decompose the molar density $h$ in the adapted coordinates as follows
\be
  h = h_s \delta(z^3 - z^3_0),
\label{8}
\ee
where $h_s(z^1,z^2)$ is a two-dimesional density on $\Sigma_z$. If we change
the adapted coordinates, we have
\[
  \tilde{h} = \frac{\partial(z^1,z^2,z^3)}{\partial(\tilde{z}^1,\tilde{z}^2,
  \tilde{z}^3)} h
  = h_s \frac{\partial(z^1,z^2)}{\partial(\tilde{z}^1,\tilde{z}^2)}
  \cdot \delta(z^3 - z^3_0)\frac{\partial z^3}{\partial \tilde{z}^3}
  = \tilde{h}_s\cdot\delta(\tilde{z}^3 - \tilde{z}^3_0),
\]
so the decomposition is independent of the choice of adapted coordinates, and
defines, in fact, a two-dimensional matter space $\Sigma_z$ with a
two-dimesional mole density $h_s$; later, we will pass to this space.

In the spacetime $M$ with coordinates $x^\mu$, the matter fields are
$z^a(x^\mu)$; the shell occupies a three-dimensional surface $\Sigma$,
which can be described by the embedding functions $x^\mu =
x^\mu(\xi^\alpha)$, $\alpha = 0,1,2$,
or by means of the equation $z^3(x^\mu) = z^3_0$.
Later, we will pass to the the matter fields $z^A = z^A(\xi^\alpha)$,
where $z^A(\xi^\alpha) = z^A(x^\mu(\xi^\alpha))$.

The gravitational field is described by the metric $g_{\mu\nu}(x)$; we
require:
\begin{description}
\item[Condition 1'] the spacetime $(M,g)$ is asymptotically flat and globally
  hyperbolic;
\item[Condition 2'] there are coordinates $x^\mu$ in a neighbourhood of each
  point of $\Sigma$ such that the metric $g_{\mu\nu}(x)$ is $C^0$ everywhere,
  piecewise $C^\infty$, so that the only discontinuity is a jump in the first
  derivatives at $\Sigma$.
\end{description}
The second derivatives of the metric will then have a delta-function
singularity at $\Sigma$ so that the Einstein's equations can be satisfied.

The total action for the system consisting of the shell and the gravitational
field can then be written in the following form
\be
  I = \frac{1}{16\pi\mbox{G}}\int_Md^4x\,\sqrt{|g|}R
  - \int_Md^4x\,\sqrt{|g|}ne(n).
\label{action}
\ee
One can use this volume form of the action to derive the equations of motion.
However, there is also a `surface form' of the action; we are going to
derive this one in the next section.

\subsection{The surface form of the action}
The coordinates satisfying Condition 2' are not uniquely determined.  We can
use this freedom for the derivation of the surface form; the tool will be the
adapted spacetime coordinates $x^\mu$, defined by the property
\[
  x^3 = z^3(x),\quad \xi^\alpha = x^\alpha|_\Sigma.
\]

Then, the induced metric $\gamma_{\alpha\beta}$ on the shell is
\be
  \gamma_{\alpha\beta} = g_{\alpha\beta}|_\Sigma
\label{9}
\ee
and its determinant $\gamma$ is related to the determinant $g$ of the
four-metric $g_{\mu\nu}$ by
\be
  \gamma = g\cdot g^{33}.
\label{10}
\ee

\subsubsection{Matter action}
Formula (\ref{1.1}) together with Eq.\ (\ref{8}) give
\[
  j^\mu = h_s\delta(z^3 - z^3_0)\epsilon^{\mu\nu\rho\sigma}z^1_\nu z^2_\rho
  z^3_\sigma.
\]
Hence, $j^3 = 0$, and as $\delta(z^3 - z^3_0) = z^3_3\delta(x^3 - x^3_0)$,
$j^\alpha$ can be written as
\be
  j^\alpha = j_s^\alpha\delta(x^3 - x^3_0).
\label{11}
\ee
The mole density $n$ can be calculated from Eq.\ (\ref{1.3}). We obtain
\[
  \sqrt{-g_{\mu\nu}j^\mu j^\nu} =
  \sqrt{-g_{\alpha\beta}j_s^\alpha j_s^\beta}\delta(x^3 - x^3_0)
  = \sqrt{-\gamma_{\alpha\beta}j_s^\alpha j_s^\beta}\delta(x^3 - x^3_0),
\]
and Eq. (\ref{10}) yields:
\[
  n = \frac{1}{\sqrt{|\gamma|}}
  \sqrt{-\gamma_{\alpha\beta}j_s^\alpha j_s^\beta}
  \sqrt{g^{33}}\delta(x^3 - x^3_0).
\]
We define the surface mole density $ n_s$ by
\be
   n_s = \frac{1}{\sqrt{|\gamma|}}
  \sqrt{-\gamma_{\alpha\beta}j_s^\alpha j_s^\beta}
\label{12}
\ee
so that
\be
  n =  n_s\sqrt{g^{33}}\delta(x^3 - x^3_0) \ ,
\label{13}
\ee
where $\sqrt{g^{33}}\delta(x^3 - x^3_0)$ is already a scalar with
respect the reparametrizations of $x^3$. For the velocity $u^\mu$, we
have the expansion
\be
  u^\mu = v^\alpha e^\mu_\alpha,
\label{14}
\ee
where
\[
  e^\mu_\alpha := \frac{\partial x^\mu}{\partial \xi^\alpha}
\]
so that
\[
  \gamma_{\alpha\beta}v^\alpha v^\beta = -1
\]
and
\[
  j_s^\alpha = \sqrt{|\gamma|} n_s v^\alpha,\quad j_s^\alpha =
  h_s\epsilon^{\alpha\beta\gamma}z^1_\beta z^2_\gamma.
\]
Then the transcription of the matter Lagrangian density is straightforward:
\[
  L_m = - \sqrt{-g}ne(n) = - \sqrt{|\gamma|} n_s e_s( n_s)\delta(x^3 - x^3_0),
\]
where $e_s( n_s)$ is the energy per mole of the shell matter. We define the
surface Lagrange density $L_s$ and the surface mass density $\rho_s$ by
\[
  L_m := L_s\delta(x^3 - x^3_0),\quad \rho_s :=  n_s e_s( n_s),
\]
so that
\be
  L_s = - \sqrt{|\gamma|}\rho_s( n_s).
\label{19.1}
\ee
If we perform the trivial integration over $x^3$ in $I_m$, the matter action
becomes a surface integral
\be
  I_m = - \int_\Sigma d^3\xi\,\sqrt{|\gamma|} n_s e_s( n_s).
\label{20.1}
\ee
This expression is invariant under the transformation of coordinates at the
shell.

The action $I_m$ can be varied with respect to the shell metric
$\gamma_{\alpha\beta}$ with the result
\[
  \delta I_m = \int_\Sigma d^3\xi\,
  \left(-\frac{1}{2}\sqrt{|\gamma|}\gamma^{\alpha\beta}\rho_s
  - \sqrt{|\gamma|}\rho_s'
  \frac{\partial n_s}{\partial\gamma_{\alpha\beta}}\right)
  \delta\gamma_{\alpha\beta}.
\]
In analogy with the formula (\ref{1.5}), we obtain
\be
  \frac{\partial n_s}{\partial\gamma_{\alpha\beta}}
  =  - \frac{1}{2} n_s(\gamma_{\alpha\beta} + v^\alpha v^\beta),
\label{28.2}
\ee
and we have
\be
  \delta I_m = \frac{1}{2}\int_\Sigma d^3\xi\,T_s^{\alpha\beta}
  \delta\gamma_{\alpha\beta}.
\label{28.3}
\ee
where $T_s^{\alpha\beta}$ is the surface stress-energy tensor,
\be
  T_s^{\alpha\beta} := \sqrt{|\gamma|}\left(\rho_sv^\alpha v^\beta
  - \sigma(\gamma_{\alpha\beta} + v^\alpha v^\beta)\right),
\label{28.4}
\ee
and
\[
  \sigma :=  n_s\rho_s' - \rho_s = -  n_s^2e_s'
\]
is the surface tension (negative two-dimensional presure). Moreover, it holds
that
\[
  T^{\mu\nu} = \frac{\partial x^\mu}{\partial \xi^\alpha}\frac{\partial
  x^\nu}{\partial \xi^\beta}
  T_s^{\alpha\beta}\sqrt{g^{33}}\delta(x^3 - x^3_0).
\]
We obtain easily relations analogous to the Eqs.\ (\ref{1.9}), (\ref{1.10}) and
(\ref{1.11}):
\be
  T^{\alpha\beta}_s(\xi) = 2\frac{\partial
  L_s}{\partial\gamma_{\alpha\beta}(\xi)}
\label{TLs}
\ee
(this is a form of Eq.\ (\ref{28.4}));
\be
  \frac{\partial L_s}{\partial z^A_\alpha}z^A_\beta
  + 2\frac{\partial L_s}{\partial\gamma_{\alpha\gamma}}\gamma_{\beta\gamma}
  = L_s\delta^\alpha_\beta,
\label{3belinf}
\ee
because $L_s$ is a three-density on $\Sigma$, and the Noether identity
\be
  T^\alpha_{s\beta} = L_s\delta^\alpha_\beta
  - \frac{\partial L_s}{\partial z^A_\alpha}z^A_\beta.
\label{Ts-can}
\ee

\subsubsection{Gravitation action}
The next task is to rewrite the shell part of the gravitational action $I_g$
in the surface form. The following lemma is vital.
\begin{lem}
In the adapted coordinates $x^\mu$ that satisfy Condition 2', the
delta-function part of the gravitational Lagrange density is given by
\be
  \frac{1}{16\pi\mbox{G}}\sqrt{|g|}R = -
  \frac{1}{8\pi\mbox{G}}\sqrt{|\gamma|}[L]\delta(x^3 - x^3_0) + \ldots,
\label{deltafL}
\ee
where the dots represent regular terms, $L =
\gamma_{\alpha\beta}L^{\alpha\beta}$, $L^{\alpha\beta}$ is the second
fundamental form of $\Sigma$ corresponding to the normal oriented outwards of
$V^-$ and the abbreviation $[f] := f_+ - f_-$ for the jump
of a quantity $f$ accross the shell is used.
\end{lem}
The proof of the lemma is relegated to the Appendix.  Eq.\ (\ref{deltafL})
implies immediately that the gravitational action can be transformed to
\be
  I_g = \frac{1}{16\pi\mbox{G}}\int_{V^+\cup V^-}d^4x\,\sqrt{|g|}R -
  \frac{1}{8\pi\mbox{G}}\int_\Sigma d^3\xi\,\sqrt{|\gamma|}[L].
\label{25.5}
\ee

Eqs.\ (\ref{20.1}) and (\ref{25.5}) give the total action in the surface
form:
\bea
  I & = & \frac{1}{16\pi\mbox{G}}\int_{V^-}d^4x\,\sqrt{|g|}R
  + \frac{1}{16\pi\mbox{G}}\int_{V^+}d^4x\,\sqrt{|g|}R \nn \\
  & - & \frac{1}{8\pi\mbox{G}}\int_\Sigma d^3\xi\,\sqrt{|\gamma|}[L]
  - \int_\Sigma d^3\xi\,\sqrt{|\gamma|} n_s e_s( n_s).
\label{Isurf}
\eea
This action functional is equivalent to that given by Eq.\ (\ref{action}), if
the coordinates satisfy Condition 2'. It has, however, two advantages in
comparison with (\ref{action}):
\ben
\item all integrands in (\ref{Isurf}) are smooth,
\item it is valid and can be used with more general coordinates, namely
  arbitrary smooth coordinates $x^\mu_\pm$ within $V^\pm$ and arbitrary
  coordinates $\xi^\alpha$ within $\Sigma$.
\een
The fields in the action
  (\ref{Isurf}) are the matter fields $z^A(y)$ on $\Sigma$ (observe that the
  fictitious field $z^3$ disappeared from the action), the gravity fields
  $g_{\mu\nu}(x)$ in $V^\pm$ and $\gamma_{\alpha\beta}(y)$ in $\Sigma$. The
  metric has to satisfy the so-called {\em continuity relations}
\be
  \gamma_{\alpha\beta}(\xi) = \left(g_{\mu\nu}^-\frac{\partial
      x^\mu_-}{\partial \xi^\alpha}\frac{\partial x^\nu_-}{\partial
      \xi^\beta}(x_-(\xi))\right)^- = \left(g_{\mu\nu}^+\frac{\partial
      x^\mu_+}{\partial \xi^\alpha}\frac{\partial x^\nu_+}{\partial
      \xi^\beta}(x_+(\xi))\right)^+,
\label{gamma-g}
\ee
where the symbols $()^\pm$ denote the limits from the volumes $V^\pm$ towards
$\Sigma$. The role of the continuity relations (\ref{gamma-g}) is to {\em
  define} the configuration space of our system similarly as a control mode
or some fall-off conditions do. The embedding functions $x^\mu_\pm(\xi)$ are
fixed; their variation is zero.

We also have to specify the integration volumes; this will be done
in analogy to Sec. \ref{sec:descr-step}: the volume $V$ is chosen to be
bounded by two Cauchy surfaces, $S_1$ and $S_2$, and by a timelike surface
$\Sigma^+$ (which will be eventually pushed to the infinity); the surface
$\Sigma$ separates $V$ in two parts, $V^\pm$, and the surfaces $S_i$ into
$S_i^\pm$; the intersections of $\Sigma$ with $S_i$ will be denoted by
$\partial\Sigma_i$ and we will assume that they together form the complete
boundary of $\Sigma$; the intersections of $\Sigma^+$ with $S_i$ will be
denoted by $\partial S_i$ and we will assume that they form the complete
boundary of $\Sigma^+$.

This form of the action will be our starting point to the derivation of the
field equations as well as the Hamiltonian formalism.

\subsection{The variational formula}
The variation of the matter part $I_m$ of the action (\ref{Isurf}), if we
calculate in the coordinates $\xi^\alpha$ that are adapted to the surfaces
$\partial \Sigma_i$ by $\xi^0 = t_i$ and $\eta^A = \xi^A$ at $\partial
\Sigma_i$, is
\beann
  \delta I_m & = & \int_\Sigma d^3\xi\,\left(\frac{\partial L_s}{\partial z^A}
  - \partial_\alpha\frac{\partial L_s}{\partial z^A_\alpha}\right)\delta z^A
  + \frac{1}{2}\int_\Sigma d^3\xi\,T^{\alpha\beta}_s\delta\gamma_{\alpha\beta}
  \\ 
  & + & \int_{\partial\Sigma_2}d^2\eta\,
  \frac{\partial L_s}{\partial z^A_0}\delta z^A
  - \int_{\partial\Sigma_1}d^2\eta\,
  \frac{\partial L_s}{\partial z^A_0}\delta z^A.
\eeann
We define the matter momenta $p_A$ by
\[
  p_A := \frac{\partial L_s}{\partial z^A_0}.
\]
As $p_A$ is a well-defined two-surface density (cf.\ the discussion below Eq.\
(\ref{momenta})), we obtain the covariant formula:
\bea
  \delta I_m & = & \int_\Sigma d^3\xi\,\left(\frac{\partial L_s}{\partial z^A}
  - \partial_\alpha\frac{\partial L_s}{\partial z^A_\alpha}\right)\delta z^A
  + \frac{1}{2}\int_\Sigma d^3\xi\,T^{\alpha\beta}_s\delta\gamma_{\alpha\beta}
\nn \\
  & + & \int_{\partial\Sigma_2}d^2\eta\,p_A \delta z^A
  - \int_{\partial\Sigma_1}d^2\eta\,p_A \delta z^A.
\label{varIms}
\eea

To calculate the variation of the gravitational part $I_g$ of the action
(\ref{Isurf}), we first rewrite the surface integral in $I_g$ with the help of
the trace part of the Eq.\ (\ref{QL}),
\[
  Q = 2\sqrt{|\gamma|}L,
\] as
\be
  - \frac{1}{8\pi\mbox{G}}\int_\Sigma d^3\xi\,\sqrt{|\gamma|}[L] =
  - \frac{1}{16\pi\mbox{G}}\int_\Sigma d^3\xi\,
  \gamma_{\alpha\beta}[Q^{\alpha\beta}],
\label{surf-int}
\ee
so that
\be
  \delta\left( - \frac{1}{8\pi\mbox{G}}\int_\Sigma d^3\xi\,\sqrt{|\gamma|}[L]
  \right) = - \frac{1}{16\pi\mbox{G}}\int_\Sigma d^3\xi\,
  (\delta\gamma_{\alpha\beta}[Q^{\alpha\beta}]
  + \gamma_{\alpha\beta}\delta[Q^{\alpha\beta}]).
\label{varIgs}
\ee
Then, we apply Lemma 1 (Eqs.\ (\ref{varIg}) and (\ref{QL})) and Eq.
(\ref{varIgs}) with the result:
\bea
  \delta I_g
  & = & - \frac{1}{16\pi\mbox{G}}\int_{V^-}d^4x\,G^{\mu\nu}\delta g_{\mu\nu}
  - \frac{1}{16\pi\mbox{G}}\int_{V^+}d^4x\,G^{\mu\nu}\delta g_{\mu\nu} \nn \\
  & - & \frac{1}{16\pi\mbox{G}}\int_\Sigma d^3\xi\,[Q^{\alpha\beta}]
  \delta\gamma_{\alpha\beta} \nn \\
  & - & \frac{1}{16\pi\mbox{G}}\int_{S_2}d^3y\,q_{kl}\delta\pi^{kl}
  + \frac{1}{16\pi\mbox{G}}\int_{S_1}d^3y\,q_{kl}\delta\pi^{kl} \nn \\
  & - & \frac{1}{8\pi\mbox{G}}\int_{\partial\Sigma_2}d^2\eta\,
  \lambda\delta[\alpha]
  + \frac{1}{8\pi\mbox{G}}\int_{\partial\Sigma_1}d^2\eta\,
  \lambda\delta[\alpha] \nn \\
  & + & \frac{1}{8\pi\mbox{G}}\int_{\partial S_2}d^2\eta\,
  \lambda\delta\alpha
  - \frac{1}{8\pi\mbox{G}}\int_{\partial S_1}d^2\eta\,
  \lambda\delta\alpha
  - \frac{1}{16\pi\mbox{G}}\int_\Sigma d^3\xi\,
  \gamma_{\alpha\beta}\delta Q^{\alpha\beta}.
\label{varIgsurf}
\eea
Here, $q_{kl}(y)$ and $\delta q_{kl}(y)$ are continuous along $S_i$, but
$\pi^{kl}(y)$ and $\delta \pi^{kl}(y)$ have a jump at $\partial\Sigma_i$.
Eqs. (\ref{varIms}) and (\ref{varIgsurf}) imply the following formula for
the total action:
\bea
  \delta I
  & = & - \frac{1}{16\pi\mbox{G}}\int_{V^-}d^4x\,G^{\mu\nu}\delta g_{\mu\nu}
  - \frac{1}{16\pi\mbox{G}}\int_{V^+}d^4x\,G^{\mu\nu}\delta g_{\mu\nu} \nn \\
  & + & \int_\Sigma d^3\xi\,\left(\frac{\partial L_s}{\partial z^A}
  - \partial_\alpha\frac{\partial L_s}{\partial z^A_\alpha}\right)\delta z^A
  + \frac{1}{16\pi\mbox{G}}\int_\Sigma d^3\xi\,
  \left(8\pi\mbox{G}T^{\alpha\beta}_s - [Q^{\alpha\beta}]\right)
  \delta\gamma_{\alpha\beta} \nn \\
  & - & \frac{1}{16\pi\mbox{G}}\int_{S_2}d^3y\,q_{kl}\delta\pi^{kl}
  + \frac{1}{16\pi\mbox{G}}\int_{S_1}d^3y\,q_{kl}\delta\pi^{kl} \nn \\
  & - & \frac{1}{8\pi\mbox{G}}\int_{\partial\Sigma_2}d^2\eta\,
  \sqrt{\lambda}\delta[\alpha]
  + \frac{1}{8\pi\mbox{G}}\int_{\partial\Sigma_1}d^2\eta\,
  \sqrt{\lambda}\delta[\alpha] \nn \\
  & + & \int_{\partial\Sigma_2}d^2\eta\,p_A\delta z^A
  - \int_{\partial\Sigma_1}d^2\eta\,p_A\delta z^A \nn \\
  & + & \frac{1}{8\pi\mbox{G}}\int_{\partial S_2}d^2\eta\,
  \sqrt{\lambda}\delta\alpha
  - \frac{1}{8\pi\mbox{G}}\int_{\partial S_1}d^2\eta\,
  \sqrt{\lambda}\delta\alpha
  - \frac{1}{16\pi\mbox{G}}\int_\Sigma d^3\xi\,
  \gamma_{\alpha\beta}\delta Q^{\alpha\beta}.
\label{varIsurf}
\eea

From the variational formula (\ref{varIsurf}), we can read off the field
equations: in $V^-$ and $V^+$,
\be
  G^{\mu\nu} = 0;
\label{einstein-s}
\ee
at $\Sigma$, we obtain
\bea
  \frac{\partial L_s}{\partial z^A}
  - \partial_\alpha\frac{\partial L_s}{\partial z^A_\alpha} & = & 0,
\label{matter-s} \\
  \mbox{}  [Q^{\alpha\beta}] & = & 8\pi\mbox{G}T^{\alpha\beta}_s.
\label{israel}
\eea
Eq.\ (\ref{israel}) is the well-known dynamical equation for thin shells
\cite{israel}; we shall refer to it as `Israel's equation'.
It may be considered as the singular part of Einstein equations,
corresponding to $\delta$-like sources. Eq.\ (\ref{matter-s}) can be
interpreted as a three-dimensional stress-energy conservation: we can
show in an analogous way as in Sec.\ \ref{sec:Tmunu} that
\[
  \nabla_\alpha T^\alpha_{s\beta} = \left(\frac{\partial L_s}{\partial z^A}
  - \partial_\alpha\frac{\partial L_s}{\partial z^A_\alpha}\right)z^A_\beta,
\]
where $\nabla_\alpha$ is a covariant derivative associated with the metric
$\gamma_{\alpha\beta}$. This identity implies that Eq.\ (\ref{matter-s}) is
equivalent to
\[
  \nabla_\alpha T^\alpha_{s\beta} = 0
\]
(which comprises only two independent equations).

Eq.\ (\ref{varIsurf}) implies a generating formula for the field equations
analogous to Eq.\ (\ref{varL}). In order to derive this formula, we first have
to introduce a foliation of the integration volume in (\ref{varIsurf}). This
is an arbitrary smooth family of spacelike surfaces $S_t$ such that $S_i =
S_{t_i}$; we allow for the surfaces $S_t$ having a cusp at $\partial\Sigma_t
= \Sigma \cap S_t$ so that the normal $n^\mu$ can have a step discontinuity
there. This leads to jumps in $\pi^{kl}$ and $\alpha$ across
$\partial\Sigma_t$.

We also have to introduce adapted cordinates $x^\mu$ so that the surfaces
$S_t$ are given by $x^0 = t$, $\Sigma$ by $x^3 = 0$ and $\Sigma^+$ by $x^3 =
r^+$; further, $y^k = x^k|_{S_t}$, $\xi^\alpha = x^\alpha|_{\Sigma}$,
$\xi^\alpha = x^\alpha|_{\Sigma^+}$, $\eta^A = x^A|_{\partial\Sigma_t}$ and
$\eta^A = x^A|_{\partial S_t}$. Observe that the full four-metric need not be
continuous across $\Sigma$ with respect to these coordinates. Observe that $t
=$ const is a continuous surface intersecting $\Sigma$ and $\Sigma^+$, and
$\partial/\partial t$ is a continuous vector field everywhere.

Eq.\ (\ref{Isurf}) and
\[
  I = \int_{t_1}^{t_2}dt\,{\cal L}
\]
imply for ${\cal L}$:
\bea
   {\cal L} & = & \frac{1}{16\pi\mbox{G}}\int_{S_t^-}d^3y\,\sqrt{|g|}R
  + \frac{1}{16\pi\mbox{G}}\int_{S_t^+}d^3y\,\sqrt{|g|}R \nn \\
  & - & \frac{1}{8\pi\mbox{G}}\int_{\partial S_t^-} d^2\eta\,
  \sqrt{|\gamma|}[L]
  - \int_{\partial S_t^-} d^2\eta\,\sqrt{|\gamma|} n_s e_s( n_s).
\label{Lsurf}
\eea
If we rewrite Eq.\ (\ref{varIsurf}) in the form
\[
  \delta I = \int_{t_1}^{t_2}dt\,\delta{\cal L},
\]
we obtain the {\em variation formula} for our system:
\bea
  \delta{\cal L}
  & = & - \frac{1}{16\pi\mbox{G}}\int_{S_t}d^3y\,
  (q_{kl}\delta\pi^{kl})^{\mbox{.}}
  - \frac{1}{8\pi\mbox{G}}\int_{\partial\Sigma_t}d^2\eta\,
  (\lambda\delta[\alpha])^{\mbox{.}} \nn \\
  & + & \int_{\partial\Sigma_t}d^2\eta\,(p_A\delta z^A)^{\mbox{.}}
  + \frac{1}{8\pi\mbox{G}}\int_{\partial S_t}d^2\eta\,
  (\lambda\delta\alpha)^{\mbox{.}} \nn \\
  & - & \frac{1}{16\pi\mbox{G}}\int_{\partial S_t}d^2\eta\,
  \gamma_{\alpha\beta}\delta Q^{\alpha\beta}.
\label{gener-s}
\eea
Performing the variation in (\ref{Lsurf}) and comparing the result with the
R.H.S. of (\ref{gener-s}) recovers the definition of momenta and the field
equations.

\subsection{The Legendre transformation}
Let us define the Hamiltonian in analogous way to Sec.
\ref{sec:legendr-gstep}:
\bea
  \check{\cal H} & = & - {\cal L}
  - \frac{1}{16\pi\mbox{G}}\int_Sd^3y\,q_{kl}\dot{\pi}^{kl}
  - \frac{1}{8\pi\mbox{G}}\int_{S\cap\Sigma}d^2\eta\,
  \sqrt{\lambda}[\dot{\alpha}] \nn \\
  & + & \frac{1}{8\pi\mbox{G}}\int_{S\cap\Sigma^+}d^2\eta\,
  \sqrt{\lambda}\dot{\alpha} + \int_{S\cap\Sigma}d^2\eta\,p_A\dot{z}^A,
\label{Hcheck}
\eea
where $[\alpha] := - \alpha_+ - \alpha_-$ at $\Sigma$, $\alpha_+$ is
defined by the normal to $\Sigma$ that is outward to $V^+$ and the future
normal to $S^-$, $\alpha_+$ is
defined by the normal to $\Sigma$ that is outward to $V^-$ and the future
normal to $S^+$.

To calculate the variation of $\check{\cal H}$, we have to regroup terms in
Eq.\ (\ref{gener-s})
\beann
  \delta{\cal L}
  & = & \frac{1}{16\pi\mbox{G}}\left[- \int_{S^-}d^3y\,
  (q_{kl}\delta\pi^{kl})^{\mbox{.}} + 2\int_{S^-\cap\Sigma}d^2\eta\,
  (\sqrt{\lambda}\delta\alpha_-)^{\mbox{.}}\right] \\
  & + & \frac{1}{16\pi\mbox{G}}\left[- \int_{S^+}d^3y\,
  (q_{kl}\delta\pi^{kl})^{\mbox{.}} + 2\int_{S^+\cap\Sigma}d^2\eta\,
  (\sqrt{\lambda}\delta\alpha_+)^{\mbox{.}} \right. \\
  & + & \left. 2\int_{S^+\cap\Sigma^+}d^2\eta\,
  (\sqrt{\lambda}\delta\alpha)^{\mbox{.}}\right]
  + \int_{S\cap\Sigma}d^2\eta\,(p_A\delta z^A)^{\mbox{.}}
  - \frac{1}{16\pi\mbox{G}}\int_{\partial S}d^2\eta\,
  \gamma_{\alpha\beta}\delta Q^{\alpha\beta}.
\eeann
Then we vary Eq.\ (\ref{Hcheck}), substitute for $\delta{\cal L}$ the
regrouped expression, and apply Lemma 1 to each of the two volumes $V^\pm$;
the result is
\bea
  \delta\check{\cal H} & = & - \frac{1}{16\pi\mbox{G}}\int_{S}d^3y\,
  (\dot{\pi}^{kl}\delta q_{kl} - \dot{q}_{kl}\delta\pi^{kl})
  + \frac{1}{16\pi\mbox{G}}\int_{S\cap\Sigma}d^2\eta\sqrt{\lambda}\,
  (\frac{\dot{\lambda}}{\lambda}\delta[\alpha] -
  \dot{[\alpha]}\frac{\delta\lambda}{\lambda})  \nn \\
  & - & \int_{S\cap\Sigma}d^2\eta\,
  (\dot{p}_A\delta z^A - \dot{z}^A\delta p_A)
  - \frac{1}{16\pi\mbox{G}}\int_{S\cap\Sigma^+}d^2\eta\,
  \sqrt{\lambda}(\frac{\dot{\lambda}}{\lambda}\delta\alpha -
  \dot{\alpha}\frac{\delta\lambda}{\lambda}) \nn \\
  & + & \frac{1}{16\pi\mbox{G}}\int_{S\cap\Sigma^+}d^2\eta\,
  \gamma_{\alpha\beta}\delta Q^{\alpha\beta}.
\label{varHcheck}
\eea
This is the Hamiltonian variation formula for the field equations.

Let us compute the value of the Hamiltonian. To this aim, we insert from
Eqs.\ (\ref{Isurf}) and (\ref{surf-int}) into Eq.\ (\ref{Hcheck}) and apply
Lemma 2. A simple calculation leads to
\bea
  \check{\cal H} & = & \frac{1}{8\pi\mbox{G}}\int_{S^-}d^3y\,G^0_0
  + \frac{1}{8\pi\mbox{G}}\int_{S^+}d^3y\,G^0_0
  + \frac{1}{8\pi\mbox{G}}\int_{S\cap\Sigma}d^2\eta\,
  [Q^0_0] \nn \\
  & - & \int_{S\cap\Sigma}d^2\eta\,T^0_{s0}
  + \frac{1}{8\pi\mbox{G}}\int_{S\cap\Sigma^+}d^2\eta\,L^0_0.
\label{valHcheck}
\eea

The term $- \frac{1}{8\pi\mbox{\footnotesize G}}[Q^0_0]$ has an interesting
interpretation. It has been obtained as the sum
\[
  - \frac{1}{8\pi\mbox{G}}[Q^0_0] = - \frac{1}{16\pi\mbox{G}}[Q] +
  \frac{1}{8\pi\mbox{G}}[L^0_0].
\]
The first summand is the delta-function term in the gravitational Lagrangian
density in the volume form,
\[
  \frac{1}{16\pi\mbox{G}}\sqrt{|g|}R,
\]
and the meaning of the second is given by the following lemma:
\begin{lem}
If the four-metric is continuous in the adapted coordinates, then we have at
$\Sigma$:
\be
  - \frac{1}{8\pi\mbox{G}}\sqrt{|g|}R^0_0 = - \frac{1}{8\pi\mbox{G}}[L^0_0]\,
  \delta(x^3 - x^3_0) + \ldots,
\label{R00}
\ee
where the dots represent regular terms.
\end{lem}
Thus, the term is the delta-function part of the expression
\[
  - \frac{1}{8\pi\mbox{G}}G^0_0
\]
(cf.\ (\ref{GR})). If the assumptions of Lemma 4 are satisfied, then the three
first integrals on the R.H.S. of (\ref{varHcheck}) can be written in volume
form just as
\[
  - \frac{1}{8\pi\mbox{G}}\int_Sd^3y\,G^0_0.
\]

The Legendre transformation at $\Sigma^+$ similar to that in Sec.\
\ref{sec:legendr-gstep} can be performed exactly as in Sec.\
\ref{sec:legendr-gstep}. The transformation to the Euler picture in the
matter part of the shell Hamiltonian (\ref{Hcheck}) is much simpler than the
analogous transformation of the step Hamiltonian (\ref{defH}), because all
formulas of Sec.\ \ref{sec:invers} remain valid, they must only be rewritten
in three spacetime and two matter space dimensions.

\subsection{The form of the Hamiltonian}
In this section, the Hamiltonian (\ref{valHcheck}) will be expressed as a
functional of the canonical variables $N$, $N_k$, $q_{kl}$, $\pi^{kl}$,
$\lambda$ and $\alpha$.

Observe that the formulas (\ref{varHcheck}) and (\ref{valHcheck}) are valid in
any coordinates that are adapted to the foliation and that make the embedding
formulas for $\Sigma$ and $\Sigma^+$ time independent. More specifically, the
coordinate $t$ must be constant along the surfaces $S$, the embedding formulas
for $\Sigma$ and $\Sigma^+$ must read
\[
  t = \xi^0,\quad y^k_\pm = y^k_\pm(\xi^K),
\]
and the embedding formulas for $S\cap\Sigma$ in $\Sigma$ is
\[
  \xi^0 = \mbox{const}, \quad \xi^K = \eta^K.
\]
Let us recall that one important point of our method is that the boundaries
are time independent in the above sense and their variations are zero.

The 2+1 decomposition of the metric $\gamma_{\alpha\beta}$ at $\Sigma$ and
$\Sigma^+$ is analogous to that of $g_{\mu\nu}$. In particular, we define the
(surface) lapse $\nu$ and the (surface) shift $\nu_K$ by
\[
  \gamma^{\alpha\beta} = \left(
  \begin{array}{ll}
    -\frac{1}{\nu^2},& \frac{\nu^L}{\nu^2}  \\
    \frac{\nu^K}{\nu^2},& \lambda^{KL} - \frac{\nu^K\nu^L}{\nu^2}
  \end{array} \right)
\]
so that
\[
  \gamma_{00} = -\nu^2 + \lambda^{KL}\nu_K\nu_L;
\]
the 2+1 deconposition of the continuity relations (\ref{gamma-g}) reads
\bea
  \nu & = & \sqrt{N^2_\pm - (N^\bot_\pm)^2},
\label{2;1} \\
  \nu_K & = & N^\pm_k e_{\pm K}^k,
\label{2;2} \\
  \lambda_{KL} & = & q^\pm_{kl} e_{\pm K}^k e_{\pm L}^l,
\label{2;3}
\eea
where
\[
  N^\bot_\pm = N^\pm_k m^k_\pm, \quad e_{\pm K}^k = \frac{\partial
  y^k_\pm}{\partial \xi^K},
\]
and $m^k_\pm$ is the unit normal vector to $\Sigma \cap S$ tangent to $S$ and
oriented from $S^-$ to $S^+$. From this definition, it follows that
\[
  \delta m^k_\pm = \frac{1}{2} m^\pm_k m^r_\pm m^s_\pm \delta q^\pm_{rs}.
\]

Using the decomposition (\ref{we00}), we can write for the integrand of the
volume terms in Eq. (\ref{valHcheck}):
\[
  G^0_0 = - \sqrt{q}\left(N\frac{G^{\bot\bot}}{\sqrt{|g|}}
  + N^k\frac{G^\bot_k}{\sqrt{|g|}} \right);
\]
observe that the R. H. S. is invariant with respect to transformations of
coordinates $y^k$. The form of $G^{\bot\bot}$ and $G^\bot_k$ is well-known
(cf.\ \cite{MTW})
\beann
  \frac{\sqrt{q}}{\sqrt{|g|}}G^{\bot\bot} & = & \frac{2\pi^{kl}\pi_{kl} -
  \pi^2}{4\sqrt{q}} - \frac{\sqrt{q}}{2} R^{(3)}, \\
  \frac{\sqrt{q}}{\sqrt{|g|}}G^\bot_k & = & - \pi^l_{k|l},
\eeann
where $R^{(3)}$ is the curvature scalar of the metric $q_{kl}$.

Within $\Sigma$ and $\Sigma^+$, an analogous decomposition yields for the
surface terms:
\[
  T^0_{s0} = - \frac{\sqrt{\lambda}}{\sqrt{|\gamma|}} (\nu T^{\bot\bot}_s +
  \nu^K T^\bot_{sK}),
\]
and
\[
  [Q^0_0] = - \frac{\sqrt{\lambda}}{\sqrt{|\gamma|}} (\nu[Q^{\bot\bot}] +
  \nu^K[Q^\bot_K]).
\]
$T^0_{s0}$ can be expressed by means of the canonical variables $z^A$, $p_A$,
$\lambda_{KL}$, $\nu$ and $\nu_K$ in a way parallel to Sec.
\ref{sec:legendre}: the formulas are independent of the dimension of
spacetime. In general, the form is only implicit. However, the dependence of
$T^0_{s0}$ on $\lambda_{KL}$, $\nu$ and $\nu_K$ can be inferred from the
relation
\be
  T^{\alpha\beta}_s =
  -2\frac{\partial(-T^0_{s0})}{\partial\gamma_{\alpha\beta}}
\label{5;1}
\ee
analogous to Eq.\ (\ref{THg}). It follows that
\be
  \frac{\partial T^0_{s0}}{\partial\nu} =
  - \frac{\sqrt{\lambda}}{\sqrt{|\gamma|}} T^{\bot\bot}_s,
\label{Tnu}
\ee
\be
  \frac{\partial T^0_{s0}}{\partial\nu^K} =
  \frac{\sqrt{\lambda}}{\sqrt{|\gamma|}} T^\bot_{sK},
\label{TnuK}
\ee
and
\be
  \frac{\partial T^0_{s0}}{\partial\lambda_{KL}} = \frac{1}{2}T^{KL}_s.
\label{Tlambda}
\ee
In particular, $\frac{\sqrt{\lambda}}{\sqrt{|\gamma|}} T^{\bot\bot}_s$ and
$\frac{\sqrt{\lambda}}{\sqrt{|\gamma|}} T^\bot_{sK}$ are both independent of
$\nu$ and $\nu_K$.

In an analogous way as in Sec.\ \ref{sec:legendre} we obtain easily
\[
  \frac{1}{n^2_s}\left(\frac{j_s^0}{\sqrt{\lambda}}\right)^2 = 1 +
  \frac{\lambda^{KL}z^A_Kz^B_Lp_Ap_B}{(j_s^0)^2\rho_s^{\prime 2}}
\]
and
\[
  T^0_{s0} = -\nu\left(\frac{\sqrt{\lambda}n_s}{\rho'_s(j^0_s)^2}
  \lambda^{KL}z^A_Kz^B_Lp_Ap_B + \sqrt{\lambda}\rho_s\right) - \nu^Kz^A_Kp_A.
\]
For a dust shell, we have
\[
  \frac{1}{n^2_s}\frac{j^0_s}{\sqrt{\lambda}} = 1 +
  \frac{\lambda^{KL}z^A_Kz^B_Lp_Ap_B}{\mu^2(j^0_s)^2}
\]
and
\[
  T^0_{s0} = -\nu\sqrt{\mu^2(j^0_s)^2 + \lambda^{KL}z^A_Kz^B_Lp_Ap_B} -
  \nu^K z^A_Kp_A.
\]
Eq.\ (\ref{Tlambda}) then yields
\[
  T_{sKL} = \nu\frac{p_Ap_Bz^A_Kz^B_L}{\sqrt{\mu^2(j^0_s)^2 +
  \lambda^{KL}z^A_Kz^B_Lp_Ap_B}}.
\]

The form of $[Q^0_0]$ can be given explicitly. Let us observe (cf.\ \cite{JK2})
that the normals $\tilde{m}$, $n$ and $m$ are related by
\beann
  \tilde{m}^\mu & = & n^\mu \sinh\alpha + m^\mu\cosh\alpha, \\
  \tilde{n}^\mu & = & n^\mu \cosh\alpha + m^\mu\sinh\alpha;
\eeann
recall that $n$ is the normal to $S$ in $M$, $m$ is the normal to $S \cap
\Sigma$ in $S$ ($m$ is orthogonal to $n$), $\tilde{n}$ is the normal to $S
\cap \Sigma$ in $\Sigma$ and $\tilde{m}$ is the normal to $\Sigma$ in $M$
($\tilde{n}$ is orthogonal to $\tilde{m}$).
A simple calculation then confirms that the corresponding second fundamental
forms  $L_{\alpha\beta}$, $K_{kl}$ and $l_{KL}$ of $\Sigma$ in $M$, $S$ in $M$
and $\Sigma \cap S$ in $S$, respectively, satisfy the relations
\[
  L_{\alpha\beta} e^\alpha_K e^\beta_L = - K_{kl} e^k_K e^l_L \sinh\alpha +
  l_{KL} \cosh\alpha,
\]
\[
  L_{\alpha\beta} \tilde{n}^\alpha e^\beta_L - K_{kl} m^k e^l_L = -\alpha_{,L}.
\]
It follows that
\[
  \frac{1}{\sqrt{|\gamma|}}Q^{\bot\bot} = \frac{1}{\sqrt{q}}\pi^{\bot\bot}
  \sinh\alpha - l \cosh\alpha,
\]
\[
  \frac{1}{\sqrt{|\gamma|}}Q^\bot_K - \frac{1}{\sqrt{q}}\pi^\bot_K =
  \alpha_{,K}.
\]
Hence,
\be
  Q^0_0 = - \nu(\sqrt{\lambda}\tilde{\pi}^{\bot\bot}\sinh\alpha -
  \sqrt{\lambda} l \cosh\alpha) - \nu^K(\sqrt{\lambda}\tilde{\pi}^\bot_K +
  \sqrt{\lambda} \alpha_{,K}),
\label{6;1}
\ee
where
\[
  l = \lambda^{KL} m_{k|l}e^k_K e^l_L = q^{kl}m_{k|l}
\]
depends only on $q_{kl}$ and its first derivatives. We use also the
abbreviations
\[
  \tilde{\pi}^{\bot\bot} = \frac{\pi^{kl}}{\sqrt{q}}m_k m_l, \quad
  \tilde{\pi}^\bot_K = \frac{\pi^{kl}}{\sqrt{q}}q_{lr}m_k e^r_K,\quad
  \tilde{\pi}_{KL} = \frac{\pi^{kl}}{\sqrt{q}}e^k_K e^l_L,
\]
where
\[
  e^k_K := \frac{\partial y^k}{\partial\eta^K}
\]
Finally, the complete Hamiltonian reads
\bea
  \check{\mathcal{H}} & = & \frac{1}{16\pi\mbox{G}} \int_{S^-}d^3y \left\{
  N\left(\frac{2\pi^{kl}\pi_{kl} - \pi^2}{2\sqrt{q}} -
  \sqrt{q}R^{(3)}\right) + N^k(-2\pi^l_{k|l})\right\} \nn \\
  & + & \frac{1}{16\pi\mbox{G}} \int_{S^+}d^3y \left\{
  N\left(\frac{2\pi^{kl}\pi_{kl} - \pi^2}{2\sqrt{q}} -
  \sqrt{q}R^{(3)}\right) + N^k(-2\pi^l_{k|l})\right\} \nn \\
  & - & \frac{1}{8\pi\mbox{G}} \int_{S\cap\Sigma}d^2\eta\sqrt{\lambda} \left(
  \nu[\tilde{\pi}^{\bot\bot}\sinh\alpha - l\cosh\alpha] +
  \nu^K[\tilde{\pi}^\bot_K + \alpha_{,K}]\right) \nn \\
  & - & \int_{S\cap\Sigma}d^2\eta\,T^0_{s0} + \frac{1}{8\pi\mbox{G}}
  \int_{S\cap\Sigma^+}d^2\eta\,L^0_0.
\label{7;1}
\eea
The surface term at $\Sigma^+$ is left unchanged; it has to be transformed
according to the control mode used and/or shifted to infinity.

The surface super-Hamiltonian ${\mathcal H}_s$ and the surface supermomentum
${\mathcal H}_{sK}$ at the shell are given by \beann {\mathcal H}_s & = &
-\frac{1}{8\pi\mbox{G}}[\tilde{\pi}^{\bot\bot}\sinh\alpha
- l\cosh\alpha] + \tilde{T}^{\bot\bot}_s, \\
{\mathcal H}_{sK} & = &-\frac{1}{8\pi\mbox{G}}[\tilde{\pi}^\bot_K +
\alpha_{,K}] + \tilde{T}^\bot_{sK}; \eeann here \beann \tilde{T}^{\bot\bot}_s
& = &
\frac{1}{\sqrt{|\gamma|}}T^{\alpha\beta}_s\tilde{n}_\alpha\tilde{n}_\beta,
\\
\tilde{T}^\bot_{sK} & = &
\frac{1}{\sqrt{|\gamma|}}T^{\alpha\beta}_s\tilde{n}_\alpha e_{\beta K}.
\eeann
The geometric meaning of the gravitational part of ${\mathcal H}_s$ and
${\mathcal H}_{sK}$ can be inferred from Eq.\ (\ref{QL}):
\beann
\frac{Q^{\bot\bot}}{\sqrt{|\gamma|}} &=& - L_{\alpha\beta}e^\alpha_Ke^\beta_L
\gamma^{KL}, \\
\frac{Q^\bot_K}{\sqrt{|\gamma|}} &=& - L_{\alpha\beta}\tilde{n}^\alpha
e^\beta_K.
\eeann
In particular, $L_{\alpha\beta}e^\alpha_Ke^\beta_L$ is the
second fundamental form of the two-surface $S \cap \Sigma$ corresponding to
the normal $\tilde{m}$ (each two-surface has two independent second
fundamental forms in the spacetime); hence, the gravity part of the surface
super-Hamiltonian is the jump in the (two-)trace of this form.

\subsection{Equations of motion}
In this subsection, we calculate the variation of the Hamiltonian (\ref{7;1})
explicitly. In this way, we can check if our method leads to the well-known
equations of motion; moreover, we can study
the structure of the canonical equations and constraints at the surface
$\Sigma$. In varying the Hamiltonian, we must carefully deal with boundary
terms.

The variation of the volume integrands can be given the form:
\[
  \delta(2G^0_0) = C_k\delta N^k + C\delta N + a^{kl}\delta q_{kl} +
  b_{kl}\delta\pi^{kl} + \sqrt{q}B^k_{|k},
\]
where
\[
  C_k = -2\pi^l_{k|l},
\]
\[
  C = \frac{1}{\sqrt{q}}(\pi^{kl}\pi_{kl} - \frac{1}{2}\pi^2) -
  \sqrt{q}R^{(3)},
\]
\beann
  a^{kl} & = & \frac{N}{\sqrt{q}}(2\pi^k_m\pi^{lm} - \pi\pi^{kl} -
  \frac{1}{2}\pi^{mn}\pi_{mn}q^{kl} \\
  & + & \frac{1}{4}\pi^2 q^{kl}) +
  N\sqrt{q}(R^{(3)\,kl} - \frac{1}{2}R^{(3)}q^{kl})
  + \sqrt{q}(N^{\ m}_{|m}q^{kl} - N^{\ kl}_{|})
  - {\mathcal L}_{\vec{N}}\pi^{kl},
\eeann
\[
  b_{kl} = \frac{N}{\sqrt{q}}(2\pi_{kl} - \pi q_{kl}) -
  {\mathcal L}_{\vec{N}}q_{kl},
\]
and
\beann
  B^r & = & - \frac{1}{\sqrt{q}}(N^k\pi^{lr} + N^l\pi^{kr} -
  N^r\pi^{kl})\delta q_{kl} - \frac{2}{\sqrt{q}}N_l\delta\pi^{lr} \\
  & - & N(q^{kl}\delta\Gamma^r_{kl} - q^{kr}\delta\Gamma^l_{kl}) +
  \frac{1}{2}N_{,s}(q^{rk}q^{sl} + q^{rl}q^{sk} - 2q^{rs}q^{kl})\delta q_{kl}.
\eeann
Here, ${\mathcal L}_{\vec{X}}$ is the Lie derivative with respect to the
vector field $\vec{X}$. The comparison with the volume term of
Eq. (\ref{varHcheck}) yields the well-known canonical form of Einstein
equations (cf.\ \cite{MTW}):
\[
  C_k = 0,\quad C = 0,
\]
\be
  \dot{\pi}^{kl} = - a^{kl}, \quad \dot{q}_{kl} = b_{kl}.
\label{8;1}
\ee
The divergence term contributes to the variation of the surface term at
$S\cap\Sigma$ by
\[
  \frac{1}{16\pi\mbox{G}}\int_{S\cap\Sigma}d^2\eta\,\sqrt{\lambda}(-B^k_+m^+_k
  + B^k_-m^-_k).
\]
The following identity can be easily derived
\be
  q^{kl}m_r(\delta\Gamma^r_{kl} - \delta^r_k\Gamma^s_{ls}) = 2\delta l +
  2l^{KL}\delta\lambda_{KL} - \lambda^{KL}(m^ke^l_L\delta
  q_{kl})_{\parallel K},
\label{9;3}
\ee
where the symbol ``$\parallel$'' denotes the covariant derivative associated
with the metric $\lambda_{KL}$ on $S\cap\Sigma$. Using Eq.\ (\ref{9;3}) and
the continuity relations (\ref{2;2}) and (\ref{2;3}), we can rewrite the
surface term in the 2+1 form
\bea
  \sqrt{\lambda}B^km_k & = & - \sqrt{\lambda}(2\tilde{\pi}^{\bot\bot}N^\bot +
  \tilde{\pi}^{\bot K}\nu_K)(m^km^l\delta q_{kl}) -
  2\sqrt{\lambda}\tilde{\pi}^{\bot\bot}\nu^K(e^k_Km^l\delta q_{kl}) \nn \\
  & - & \sqrt{\lambda}(\tilde{\pi}^{\bot L}\nu^K + \tilde{\pi}^{\bot K}\nu^L -
  N^\bot\tilde{\pi}^{KL} + \tilde{\pi}^{\bot\bot}N^\bot\lambda^{KL}
  + \tilde{\pi}^{\bot M}\nu_M\lambda^{KL}  \nn \\
  & + & N_{,k}m^k\lambda^{KL} - Nl^{KL})\delta\lambda_{KL} \nn \\
  & + & 2\sqrt{\lambda}N\delta l - 2\sqrt{\lambda}(N^\bot m_km_l +
  \nu_Ke^K_km_l)\delta\tilde{\pi}^{kl}.
\label{9;1}
\eea

The variation of the surface term $[Q^0_0]$ can be written in the following
way
\be
  \delta Q^0_0 = \delta\nu \left(
  -\frac{\sqrt{\lambda}}{\sqrt{|\gamma|}}Q^{\bot\bot} \right) - \delta\nu^K
  \left( \frac{\sqrt{\lambda}}{\sqrt{|\gamma|}}Q^\bot_K\right) + \mbox{Rest}.
\label{9;2}
\ee
Similarly, using Eqs.\ (\ref{Tnu}), (\ref{TnuK}) and (\ref{Tlambda}), we
obtain that
\bea
  \delta T^0_{s0} & = & \delta\nu \left(
  -\frac{\sqrt{\lambda}}{\sqrt{|\gamma|}}T_s^{\bot\bot} \right) - \delta\nu^K
  \left( \frac{\sqrt{\lambda}}{\sqrt{|\gamma|}}T^\bot_{sK} \right) +
  \frac{1}{2} T^{KL}_s \delta\lambda_{KL} \nn \\
  & + & \left( \frac{\partial
  T^0_{s0}}{\partial z^A} - \frac{\partial}{\partial\eta^M} \frac{\partial
  T^0_{s0}}{\partial z^A_M} \right)\delta z^A + \frac{\partial
  T^0_{s0}}{\partial p_A} \delta p_A.
\label{10;1}
\eea
Comparing the first two terms in Eqs. (\ref{9;1}) and (\ref{10;1}), we obtain
the first three Israel's equations:
\be
  [Q^{\bot\bot}] = 8\pi\mbox{G} T^{\bot\bot}_s, \quad
  [Q^\bot_K] = 8\pi\mbox{G} T^\bot_{sK}.
\label{10;2}
\ee
The last two terms in Eq.\ (\ref{10;1}), if compared with the corresponding
surface term in Eq.\ (\ref{varHcheck}) yield the dynamical equations for
matter inside the three-dimensional spacetime of the shell surface $\Sigma$:
\be
  \frac{\partial T^0_{s0}}{\partial z^A}
  - \frac{\partial}{\partial\eta^M} \frac{\partial T^0_{s0}}{\partial z^A_M}
  = \dot{p}_A,
\label{10;3}
\ee
\be
  \frac{\partial T^0_{s0}}{\partial p_A} \delta p_A = - \dot{z}^A.
\label{10;4}
\ee
What remains from Eq.\ (\ref{varHcheck}) can be written as follows
\be
  [2\mbox{Rest} - \sqrt{\lambda}B^km_k] = \left(8\pi\mbox{G}T^{KL}_s -
  \sqrt{\lambda}[\dot{\alpha}]\lambda^{KL}\right) \delta\lambda_{KL}
  + \sqrt{\lambda}\lambda^{KL}\dot{\lambda}_{KL}\delta[\alpha].
\label{11;1}
\ee
A somewhat lenghty calculation starting with Eqs.\ (\ref{9;1}) and (\ref{6;1})
gives
\beann
  \lefteqn{2\mbox{Rest} - \sqrt{\lambda}B^km_k = } \\
  && - 2\sqrt{\lambda}\tilde{\pi}^{\bot\bot}(\nu\sinh\alpha - N^\bot)
  m^km^l\delta q_{kl} - \sqrt{\lambda}\{\tilde{\pi}^{\bot\bot}\lambda^{KL}
  (\nu\sinh\alpha - N^\bot) \\
  && + \tilde{\pi}^{KL}N^\bot + l^{KL}N - \nu l \lambda^{KL}\cosh\alpha +
  \nu^M\lambda^{KL}\alpha_{,M} - N_{,k}m^k\lambda^{KL}\}\delta\lambda_{KL} \\
  && + 2\sqrt{\lambda}(\nu\cosh\alpha - N)\delta l -
  2\sqrt{\lambda}(\nu\sinh\alpha - N^\bot)m_km_l\delta\tilde{\pi}^{kl} \\
  && + 2\sqrt{\lambda}(-\nu\tilde{\pi}^{\bot\bot}\cosh\alpha + \nu
  l\sinh\alpha + \nu^K_{\parallel K})\delta\alpha.
\eeann
Substituting this into Eq.\ (\ref{11;1}), we obtain immediately
\be
  \nu\sinh\alpha_\pm = N^\bot_\pm, \quad  \nu\cosh\alpha_\pm = N_\pm,
\label{12;1}
\ee
\be
  [-\tilde{\pi}^{\bot\bot}\cosh\alpha + l\sinh\alpha] = 0.
\label{12;2}
\ee
The remaining equations, simplified by (\ref{12;1}) and (\ref{12;2}), read
\be
  \sqrt{\lambda}[-\tilde{\pi}^{KL}N^\bot - l^{KL}N + Nl \lambda^{KL} +
  N_{,k}m^k\lambda^{KL} + \nu^M\lambda^{KL}\alpha_{,M}
  + \dot\alpha\lambda^{KL}] = 8\pi\mbox{G}T^{KL}_s,
\label{12;3}
\ee
and
\be
  \dot{\lambda} = -2\lambda(N\tilde{\pi}^{\bot\bot} - N^\bot l -
  \nu^K_{\parallel K}).
\label{12;4}
\ee

From the point of view of physical (or geometrical) content, Eq.\
(\ref{12;1}) just reproduces the definition of $\alpha$ and it is
compatible with the continuity relation (\ref{2;1}). The three Eqs.\
(\ref{12;3}) are equivalent to the remaining three Israel's equations.
Finally, Eqs.\ (\ref{12;2}) and (\ref{12;4}) follow from the continuity
relations (\ref{2;1})--(\ref{2;3}) and the equation of motion
(\ref{8;1}). Indeed, taking limit of the second Eq.\ (\ref{8;1}) from
both sides towards the shell and projecting the result by $e^k_Ke^l_L$,
we obtain
\be
  \dot{\lambda}_{KL} = \left( 2N(\tilde{\pi}_{KL} -
  \frac{1}{2}\tilde{\pi}\lambda_{KL}) + N_{K\parallel L} + N_{L\parallel K} +
  2 l_{KL}N^\bot \right)_\pm;
\label{13;3}
\ee
the expression in the brackets on the R.H.S. must be continuous, hence
\be
  [-N(\tilde{\pi}\lambda_{KL} - 2\tilde{\pi}_{KL}) + 2 l_{KL}N^\bot] = 0.
\label{13;4}
\ee
The trace of Eq.\ (\ref{13;4}) is Eq.\ (\ref{12;2}) and the trace of Eq.\
(\ref{13;3}) is Eq.\ (\ref{12;4}).

Eqs.\ (\ref{10;2}), (\ref{13;4}) and the tracefree part
of Eq.\ (\ref{12;3}) can be written in a more symmetric form
\bea
  [-K_{KL}\sinh\alpha + l_{KL}\cosh\alpha] & = &
  - 8\pi\mbox{G}(\tilde{T}_{sKL} - \frac{1}{2}\tilde{T}\lambda_{KL}),
\label{constr1} \\ \mbox{}
  [K_{KL}\cosh\alpha - l_{KL}\sinh\alpha] & = & 0,
\label{constr2} \\ \mbox{}
  [\tilde{\pi}^\bot_K + \alpha_{,K}] = - 8\pi\mbox{G}\tilde{T}^\bot_{sK},
\label{constr3}
\eea
where $K_{KL} = -\tilde{\pi}_{KL} + (1/2)\tilde{\pi}\lambda_{KL}$. They
give the jumps of the two independent second fundamental forms of the shell
2-surface in the spacetime, one corresponding to the normal $\tilde{n}$ in the
direction of the shell motion (continuous), the other to $\tilde{m}$, which is
perpendicular to the direction of motion.

From the point of view of the theory of constraint systems, the Eqs.\
(\ref{10;2}), (\ref{12;2}), and the tracefree part of Eq.\ (\ref{12;3})
are constraints. The trace of Eq.\ (\ref{12;3}),
\be
  [\dot{\alpha}] = 4\pi\mbox{G}\nu\tilde{T}^{KL}_s\lambda_{KL} -
  \frac{1}{2}\left[- N^\bot\tilde{\pi}^{KL}\lambda_{KL} + Nl + 2N_{,k}m^k +
  2\nu^K\alpha_{,K}\right],
\label{13;2}
\ee
Eq.\ (\ref{12;4}) and Eqs.\ (\ref{10;3}), (\ref{10;4}) are canonical
equations. Finally, Eqs. (\ref{2;1})--(\ref{2;3}) and (\ref{12;1}) are
defining equations of the Hamiltonian system, analogous to fall-off conditions
or control conditions.

It seems that some of the constraints are second class. For example, Eq.\
(\ref{12;2}) follows from the variation with respect to $\bar{\alpha} =
(1/2)(\alpha_+ + \alpha_-)$, which is a Lagrange multiplier. Eq.\ (\ref{12;2})
contains this Lagrange multiplier; thus, its Poisson bracket with
$\pi_{\bar{\alpha}}$, which is the momentum conjugate to $\bar{\alpha}$, and
which is also constrained to vanish, is not zero (if we extend the system by
this momentum).

Another important observation is that the L. H. S. of Eq.\ (\ref{12;2}) can be
smeared only by a function of two variables, because the domain of definition
of the L. H. S. is the shell surface. On the other hand, Eq.\ (\ref{12;2})
contains so-called volume quantities, namely $\pi^{kl}$ and $l^{KL}$; a
derivative with respect to these variables and the Poisson brackets of these
variables result in three-dimensional $\delta$-functions. Thus, the L. H. S.
of Eq,\ (\ref{12;2}) {\em cannot} be smeared so that it becomes a
differentiable function on the phase space. We call such constraints {\em
  singular}. The best way of tackling this constraint may be to solve it for
$\bar{\alpha}$ and insert the solution back into the action (cf.\ \cite{H-T}).
A similar procedure exists hopefully for the two constraints which result from
the tracefree part of Eq.\ (\ref{12;3}):
\beann
  \lefteqn{8\pi\mbox{G}\left(\tilde{T}^{KL}_s -
  \frac{1}{2}\tilde{T}^{MN}_s\lambda_{MN}\lambda^{KL}\right) = } \\
  && \left[-\left(\tilde{\pi}^{KL} -
  \frac{1}{2}\tilde{\pi}^{MN}\lambda_{MN}\lambda^{KL}\right) \cosh\alpha -
  \left(l^{KL} - \frac{1}{2}l^{MN}\lambda_{MN}\lambda^{KL}\right)
  \sinh\alpha\right];
\eeann
they together with Eq.\ (\ref{12;2}) exhaust the singular constraints of our
system.  It is interesting to observe that the total Hamiltonian {\em is} a
differentiable function. Although the surface integrals in the Hamiltonian
represent two-dimensional smearing of some volume quantities, the presence of
volume terms and the continuity relations between the surface ($\nu$, $\nu^K$)
and volume ($N$, $N^k$) smearing functions guarantee an effectively
three-dimensional smearing of all volume quantities.

These difficulties and the related problem of the Bergmann-Dirac analysis
\cite{EN} seem to be non-trivial; we will try to tackle them in a future
paper.

\subsection*{Acknowledgements}
The authors are indebted to G.~Lavrelashvili for carefully reading the
manuscript and for checking some equations. P.H. thanks to the
Max-Planck-Institut for Gravitationsphysik, Potsdam, where some
calculations have been performed, for the nice hospitality and support,
to the Tomalla Foundation, Zurich and to the Swiss Nationalfonds for a
partial support.
J.K. thanks to the Tomalla Foundation, Zurich and to the Polish National
Commitee for Science and Research (KBN) for a partial support.

\appendix

\section{Proof of Lemma 3}
Let us calculate the delta-function terms in the expression $L_g =
h^{\mu\nu}R_{\mu\nu}$, where $h^{\mu\nu}$ is defined by
\be
  h^{\mu\nu} := \frac{1}{16\pi\mbox{G}}|g|^{1/2}g^{\mu\nu}.
\label{hmunu}
\ee
For this aim, we have to isolate the second derivative terms. A
simple calculation gives
\[
  h^{\mu\nu}R_{\mu\nu}
  = \partial_\rho(h^{\mu\nu}A^\rho_{\mu\nu}) + \ldots,
\]
where $A^\rho_{\mu\nu}$ is defined by
\be
  A^\lambda_{\mu\nu} := \Gamma^\lambda_{\mu\nu} -
  \delta^\lambda_{(\mu}\Gamma^\kappa_{\nu)\kappa}.
\label{A}
\ee
and the dots represent regular terms. Thus, the delta-function term in
$h^{\mu\nu}R_{\mu\nu}$ is given by
\[
  h^{\mu\nu}[A^3_{\mu\nu}]\delta(x^3 - x^3_0).
\]
Thus, we are to prove the following identity
\be
  h^{\mu\nu}[A^3_{\mu\nu}] = - \frac{1}{8\pi\mbox{G}}\sqrt{|\gamma|}[L],
\label{25.4}
\ee
where $L_{\alpha\beta}$ is the second fundamental form of $\Sigma$ and $L =
\gamma^{\alpha\beta}L_{\alpha\beta}$. Observe that the R.H.S. of (\ref{25.4})
is written in a thre-covariant form.  From the definition of (\ref{A}) of the
quantity $A^\rho_{\mu\nu}$, it follows that
\bea
  A^3_{33} & = & - \Gamma^\alpha_{3\alpha},
\label{22.4} \\
  A^3_{\alpha 3} & = & \frac{1}{2}(\Gamma^3_{3\alpha}
  - \Gamma^\beta_{\beta\alpha}),
\label{22.5} \\
  A^3_{\alpha\beta} & = & \Gamma^3_{\alpha\beta}.
\label{22.6}
\eea
Let $m_\mu$ denote the unit normal vector to $\Sigma$ oriented in the
direction of inceasing $x^3$, that is outward with respect to $V^-$:
\be
  m_\mu = \frac{1}{\sqrt{g^{33}}}\delta^3_\mu.
\label{23.1}
\ee
Then,
\be
  L_{\alpha\beta} = m_{\alpha;\beta} = -
  \frac{\Gamma^3_{\alpha\beta}}{\sqrt{g^{33}}},
\label{23.2}
\ee
so
\be
  A^3_{\alpha\beta} = - \sqrt{g^{33}}L_{\alpha\beta}.
\label{23.3}
\ee
$h^{\mu\nu}$ is a tensor density, hence
\[
  \nabla_\kappa h^{\mu\nu} = h^{\mu\nu}_{,\kappa}
  + \Gamma^\mu_{\kappa\lambda}h^{\lambda\nu}
  + \Gamma^\nu_{\kappa\lambda}h^{\mu\lambda}
  - \Gamma^\lambda_{\kappa\lambda}h^{\mu\nu},
\]
but it is a tensor density formed from the components of the metric tensor,
thus $\nabla_\kappa h^{\mu\nu} = 0$ for all
$\kappa,\mu,\nu$. Setting $\mu = 3$, $\nu = \kappa = \alpha$ in this equation
and using Eqs.\ (\ref{22.4}), (\ref{23.2}) and (\ref{23.3}), we obtain that
\be
  A^3_{33} = \frac{h^{3\alpha}_{,\alpha}
  - \sqrt{g^{33}}L_{\alpha\beta}h^{\alpha\beta}}
  {h^{33}}.
\label{23.4}
\ee
Similarly
\[
  A^3_{\alpha 3} = - \frac{h^{33}_{,\alpha}
  - 2\sqrt{g^{33}}L_{\alpha\beta}h^{3\beta}}
  {2h^{33}}.
\]
The metric $g_{\mu\nu}(x)$ is continuous across the shell, so
will be $h^{\mu\nu}$ and the tangential derivatives of
$h^{\mu\nu}$. It follows that
\be
  [A^3_{33}]
  = - \frac{g^{\alpha\beta}}{\sqrt{g^{33}}}[L_{\alpha\beta}],
\label{24.1}
\ee
\be
  [A^3_{\alpha 3}]
  = \frac{g^{3\beta}}{\sqrt{g^{33}}}[L_{\alpha\beta}],
\label{24.2}
\ee
where we also have substituted for $h^{\mu\nu}$ from (\ref{hmunu}). Moreover,
using Eq.\ (\ref{23.3}), we have that
\be
  [A^3_{\alpha\beta}]
  = - \frac{g^{33}}{\sqrt{g^{33}}}[L_{\alpha\beta}].
\label{24.3}
\ee
The Eqs.\ (\ref{24.1}), (\ref{24.2}) and (\ref{24.3}) imply
\[
  h^{\mu\nu}[A^3_{\mu\nu}] = - \frac{1}{8\pi\mbox{G}}\sqrt{|g|g^{33}}
  \left(g^{\alpha\beta} - \frac{g^{3\alpha}g^{3\beta}}{g^{33}}\right)
  [L_{\alpha\beta}].
\]
However, the following well-known relation holds:
\[
  \gamma^{\alpha\beta} = g^{\alpha\beta} -
  \frac{g^{3\alpha}g^{3\beta}}{g^{33}},
\]
from which, if Eq.\ (\ref{10}) is used,
the identity (\ref{25.4}) follows immediately.

\section{Proof of Lemma 4}
If we rewrite $R^0_0$ in terms of the connection,
\[
  R^0_0 = g^{0\mu}\left(\partial_\rho\Gamma^\rho_{\mu 0} -
  \partial_0\Gamma^\rho_{\mu\rho} + \ldots\right),
\]
where the dots represent terms that do not contain second derivatives of the
metric, we obtain immediately that
\be
  R^0_0 = g^{0\mu}[\Gamma^3_{\mu 0}]\delta(x^3 - x^3_0) + \ldots.
\label{deltaR00A}
\ee
We easily find:
\[
  g^{0\mu}[\Gamma^3_{0\mu}] = -g^{33}\gamma^{0\alpha}[g_{0\alpha,3}].
\]
The following two equations are easily verified:
\beann
  g_{\alpha\beta,3} & = & \frac{1}{g^{33}}L_{\alpha\beta}, \\
  L_{\alpha\beta} & = &
  \frac{1}{|\gamma|}\left(\frac{1}{2}Q\gamma_{\alpha\beta} -
  Q_{\alpha\beta}\right).
\eeann
Then, some computation leads to
\[
  g^{0\mu}[\Gamma^3_{0\mu}] = \frac{\sqrt{g^{33}}}{\sqrt{|\gamma|}}
  ([Q^0_0] - \frac{1}{2}[Q]).
\]
Finally, using Eqs.\ (\ref{10}) and (\ref{deltaR00A}), we obtain Eq.\
(\ref{R00}) immediately.

\end{document}